\documentclass[fleqn,usenatbib]{mnras}

\usepackage{newtxtext,newtxmath}

\usepackage[T1]{fontenc}

\DeclareRobustCommand{\VAN}[3]{#2}
\let\VANthebibliography\thebibliography
\def\thebibliography{\DeclareRobustCommand{\VAN}[3]{##3}\VANthebibliography}

\usepackage{graphicx}	% Including figure files
\usepackage{amsmath}	% Advanced maths commands
\usepackage{amssymb}	% Extra maths symbols

\newcommand{\abcde}{$\alpha \beta \gamma \delta \epsilon$ }
\newcommand{\abce}{$\alpha \beta \gamma \epsilon$ }
\newcommand{\abcep}{$\alpha \beta \gamma \epsilon-$pure }
\title[Cepheid metallicity in the Leavitt law (C-MetaLL) survey – III.]{Cepheid metallicity in the Leavitt law (C-MetaLL) survey – III.\\ Simultaneous derivation of the {\it Gaia} parallax offset and Period-Luminosity-Metallicity coefficients.}

\author[R. Molinaro et al.]{
R. Molinaro$^{1}$\thanks{E-mail: roberto.molinaro@inaf.it},
V. Ripepi$^{1}$,
M. Marconi$^{1}$,
M. Romaniello$^{2}$,
G. Catanzaro$^{3}$,
F. Cusano$^{4}$,
G. De Somma$^{1,5}$,
\newauthor
I. Musella$^{1}$,
J. Storm$^{6}$,
E. Trentin$^{1,6,7}$
\\
$^{1}$INAF-OACN Osservatorio Astronomico di Capodimonte, salita Moiariello 16, Napoli (ITALY)\\
$^{2}$ European Southern Observatory (ESO), Karl-Schwarzschild-Str., D-85748 Garching, Germany\\
$^{3}$ INAF-Osservatorio Astrofisico di Catania, via Santa Sofia 78, 95125, Cataina, Italy\\
$^{4}$ INAF-Osservatorio di Astrofisica e Scienza dello Spazio, Via Gobetti 93/3, I-40129 Bologna, Italy\\
$^{5}$ Istituto Nazionale di Fisica Nucleare (INFN)-Sez. di Napoli, via Cinthia 80126 Napoli, Italy\\
$^{6}$ Leibniz Institut für Astrophysik Potsdam (AIP), An der Sternwarte 16, D-14482 Potsdam, Germany\\
$^{7}$ Institut f\"ur Physik und Astronomie,
Univ. Potsdam, Karl-Liebknecht-Strasse 24/25, D-14476 Potsdam, Germany
}

\date{Accepted XXX. Received YYY; in original form ZZZ}

\pubyear{2015}

\begin{document}
\label{firstpage}
\pagerange{\pageref{firstpage}--\pageref{lastpage}}
\maketitle

% Abstract of the paper
\begin{abstract}
Classical Cepheids (DCEPs) are the most important standard candles in the extra-galactic distance scale thanks to the Period-Luminosity ($\rm PL$), Period-Luminosity-Color ($\rm PLC$) and Period-Wesenheit ($\rm PW$) relations that hold for these objects. The advent of the {\it Gaia} mission, and in particular the Early Data Release 3 (EDR3) provided accurate parallaxes to calibrate these relations. In order to fully exploit {\it Gaia} measurements, the zero point (ZP) of {\it Gaia} parallaxes should be determined with an accuracy of a few $\rm \mu as$. The individual ZP corrections provided by the {\it Gaia} team depend on the magnitude and the position on the sky of the target.
In this paper, we use an implicit method that relies on the Cepheid $\rm PL$ and $\rm PW$ relations to evaluate the ensemble {\it Gaia} parallax zero point. 
The best inferred estimation of the offset value needed to additionally correct (after the {\it Gaia} team correction) the {\it Gaia} parallaxes of the present DCEP sample, amounts to $\rm -22\pm 4\, \mu as$. This value is in agreement with the most recent literature values and confirms that the correction proposed by the {\it Gaia} team over-corrected the  parallaxes.\\
As a further application of our results, we derive an estimate of the Large Magellanic Cloud distance ($\rm \mu_0=18.49\pm 0.06\, mag$), in very good agreement with the currently accepted value obtained through geometric methods.
\end{abstract}

% Select between one and six entries from the list of approved keywords.
% Don't make up new ones.

%%%%%%%%%%%%%%%%%%%%%%%%%%%%%%%%%%%%%%%%%%%%%%%%%%

%%%%%%%%%%%%%%%%% BODY OF PAPER %%%%%%%%%%%%%%%%%%

\section{Introduction}

Classical Cepheids (DCEPs) are young (50-500 Myr) stellar pulsators crossing the instability strip during the central helium-burning phase. They are the most important standard candles in the extra-galactic distance scale thanks to their Period-Luminosity ($\rm PL$), Period-Luminosity-Color ($\rm PLC$) and Period-Wesenheit ($\rm PW$) relations \citep[e.g.][]{Leavitt1912,Madore1982,Caputo2000,Riess2016}. 
Indeed, these relations allow us to calibrate the secondary distance indicators such as Supernovae Ia, and eventually, to estimate the value of the Hubble constant ($\rm H_0$) \citep[e.g.][]{Riess2016}. 
However, it is now well known \citep[e.g.][and references therein]{Riess2022} that there exists a discrepancy at the level of 5 $\rm \sigma$ between the value of $\rm H_0$ as measured with the  Cepheid-based and SNIa-based extra-galactic distance scale and that estimated from the early universe from the cosmic microwave background and the $\rm \Lambda$CDM (Cold Dark Matter) theory \citep[][]{Planck2020}. 
In spite of many observational and theoretical efforts, the causes of this discrepancy are still unknown. In this context, it is important to investigate in detail all the possible causes of systematic errors in the methods adopted to estimate $\rm H_0$. Concerning the cosmic distance scale, one of the possible sources of systematic error is the dependence of  $\rm PL$, $\rm PLC$ and $\rm PW$ relations on metallicity, which is still not well constrained \citep[see][]{and16, Gieren2018,gro18,rip19,bre21, bre22,rie21, Riess2022, rip21,des22,rip22,tre22}. 
To tackle this problem, we started a project dubbed C-MetaLL (Cepheids - Metallicity in the Leavitt Law, see \citet[][hereafter R21]{rip21} for full details on the project) with the aim of obtaining new high-resolution spectroscopy and homogeneous photometry both in the optical and in the near-infrared (NIR) for a significant sample (i.e. $>$ 250 objects) of DCEPs spanning a wide range of metallicity values (e.g. $-1.0<$[Fe/H]$<+0.5$). These data, in conjunction with the parallaxes from the {\it Gaia} mission \citep[][]{GaiaPrusti2016} will allow us to derive accurate $\rm PLZ$ and $\rm PWZ$ relations (Z stands for metallicity). 

However, it is well known that {\it Gaia} parallaxes are affected by an offset and their values need to be corrected before any scientific use \citep{GaiaBrown2018, lur18}. This offset, found by studying the parallax values for a sample of half a million quasars, assumed to have zero parallax, was found to depend on both the colour and the magnitude of the source, as well as on its position in the sky. In the Data Release 2 (DR2), the offset was found to be at least $\rm 29\,\mu$as or larger \citep[see e.g.][]{Arenou2018,Lindegren2018,Leung2019,rip20}. The publication of Early Data Release 3  \citep[EDR3][]{GaiaBrown2021} was accompanied by a recipe to calculate the individual offset values to be used for each star, depending on its ecliptic latitude, magnitude and colour \citep[][hereafter L21]{lin21}. However, subsequent studies found that the L21 method actually over-corrected the EDR3 parallaxes, with the need for an additional offset to obtain accurate parallaxes \citep[see e.g.][]{bha21,fab21,gil21,hua21,ren21,rie21,sta21,vas21,zin21,cru22,wan22}. 

In the first work of this series, R21 checked their PLZ relations, derived for various band combinations, against the accurate geometric distance of the (Large Magellanic Cloud) LMC \citep{pie19}. In their analysis the {\it Gaia} EDR3 parallaxes were corrected according to both the L21 recipe and including the offset determined by \citet{rie21}, but no final conclusion was inferred about the parallax offset value. Anyway, they found an interesting correlation of the parallax shift value with the metallicity coefficient of the $\rm PLZ$ and $\rm PWZ$ relations. In particular they found that the two parameters are degenerate: the effect of changing the parallax offset can be mimicked by varying the metallicity term in the $\rm PLZ$ and $\rm PWZ$ relations.

In the context of the C-MetaLL program, this paper aims to simultaneously determine the PLZ coefficients and  the {\it Gaia} parallax offset, by applying the fitting technique described in \cite{lay19} to a sample of RR Lyrae stars and successfully applied to EDR3 parallaxes by \citep[][]{gil21}.

The paper is structured as follows: the fitting technique together with the Monte Carlo simulations are introduced in  Sect.~\ref{sec-theTechnique}; the results of the applied fit, together with an application consisting in estimating the LMC distance are described in  Sect.~\ref{sec-theFitResults}; a discussion including the comparison with recent literature results is contained in Sect.\ref{sec-discussion}; finally Sect.\ref{sec-conclusions} contains our  conclusions.

\section{The fitting procedure}
\label{sec-theTechnique}

Before describing the procedure adopted to obtain the $\rm PLZ$ and $\rm PWZ$ relations with the implicit method, we briefly illustrate the observational sample used in this work.

\subsection{The observational sample}

We adopted the DCEPs sample devised in R21, which is composed of 443 stars\footnote{After sample cleaning as detailed in R21}, subdivided in 358 fundamental, 63 first overtone and 22 multi-mode\footnote{For these objects we used the longest period of pulsation} pulsators (DCEP\_F, DCEP\_1O, and DCEP\_MULTI, respectively). The catalogue includes the iron abundance ([Fe/H]) value based on high-resolution spectroscopy and photometry in the $\rm V,\,I,\,J,\,H,\,K_s$ bands\footnote{We included also the F555W, F814W and F160W Hubble Space Telescope (HST) bands obtained from ground based V, I, H bands by using the transformations by \citet[][]{bre20}}. Periods are taken from the literature or redetermined in R21, while the parallaxes with relative errors are from the {\it Gaia} EDR3 catalogue.

\subsection{The method}

To preserve the Gaussian properties of the parallax errors and to avoid introducing biases by neglecting negative parallaxes, we adopted the  Astrometric Based Luminosity (ABL) formalism \citep{fea97, are99}. Specifically, we can write the $\rm PLZ$ or $\rm PWZ$ relations as follows:
\begin{equation}
\rm ABL=\varpi 10^{0.2 m -2}=10^{0.2(\alpha+(\beta+\delta [Fe/H])(\log P-\log P_0) +\gamma [Fe/H])} \label{eq:abl}
\end{equation}
where $\varpi$ is the parallax, $\rm m$ is the observed dereddened apparent magnitude or the reddening free Wesenheit\footnote{Given a generic photometric band $\rm X$ and a generic color $\rm Y-Z$, the Wesenheit pseudo-magnitude is defined by $\rm W(X,Y-Z) = X +\lambda\cdot (Y-Z)$, where the coefficient $\lambda$ is equal to the total-to-selective extinction ratio $\rm R_X=\frac{A_X}{E(Y-Z)}$. The obtained quantity is reddening-free by definition.} pseudo-magnitude \citep[][]{Madore1982}, $\rm P$ is the period (in days) and  $\rm [Fe/H]$ is the iron abundance. The pivot value $\rm logP_0$ is also considered in the equation to reduce the correlation among fitted coefficients.

The idea consists in fitting simultaneously the coefficients of equation \ref{eq:abl} and the {\it Gaia} parallax offset. To this aim the equation \ref{eq:abl} can be written using an implicit form and introducing the parallax offset as a further parameter of the fit \citep[][]{lay19}:

\begin{equation}
\rm f =  10^{0.2(\alpha+(\beta+\delta [Fe/H])(\log P-\log P_0) +\gamma [Fe/H])} - (\varpi + \epsilon) 10^{0.2 m -2} =0, \label{eq:abl-implicit}  
\end{equation}

\noindent
where $\epsilon$ represents the parallax offset.

To face the problem of solving this implicit equation, we have used the open-source FORTRAN package ODRPACK95 \citep{zwo07}. This software (hereafter ODR) allows us to solve weighted fitting problems when errors have to be considered on all variables. In these cases, the software minimizes the orthogonal distance of the input data from the fitted line and/or surface. 
Moreover, it allows parameter estimation also in the case of implicit fit, as the one considered in this paper (Eq.~\ref{eq:abl-implicit}). The base algorithm consists of the Levenberg-Marquardt method \citep{lev44, mar63} and is also able to calculate the input function derivatives by using the finite difference method if the Jacobian matrix is not provided by the user \citep[][and references therein]{zwo07, bog89}. In our case, we provided both the fitted function and its Jacobian matrix, but we verified that the results do not change by using the internal derivative method.

According to \citet{zwo07}, fixing the weights when all variables are affected by errors, is not straightforward. The values of the weights should compensate for precision differences among the variables and their spread over intervals  covering different orders of magnitudes. As a consequence, wrong weight values can affect the fit results providing unreliable coefficients. Therefore, we performed both the weighted and the unweighted analyses in order to verify if they return consistent results. Moreover, we first carried out a large set of Monte Carlo simulations to test how the fitting routine is able to recover the $\rm PLZ$  or $\rm PWZ$ coefficients, assuming known input relations.

In the following, we apply the whole procedure to the observational sample presented above, considering both the F-mode DCEPs only and the global F+1O DCEPs sample. In the latter case, the 1O pulsators' periods were  fundamentalised according to the recipe by \citet{fea97}. The $\rm PLZ$ and  $\rm PWZ$ relations are calculated for a variety of bands and colours, specifically: $\rm K_s$, $\rm W_{V,V-K_s}$, $\rm W_{V,J-K_s}$, $\rm W_{H,V-I}$ and $\rm W^{HST}_{H,V-I}$\footnote{To transform Johnson-Cousins-2MASS ground-based $\rm H,\,V,\,I$ photometry into the HST correspondent F160W,\,F555W,\,F814W, we adopted the transformations provided by \citet{bre20}.}. To calculate the Wesenheit magnitudes we adopted the coefficients for the color terms listed in the Tab.~\ref{tab-wesCoeff}. 

\begin{table}
    \centering
    \begin{tabular}{l|c}
    \hline
    \hline
    Bands&$\lambda$\\
    \hline
    $\rm W_{V,V-K_s}$ & -0.690\\
    $\rm W_{V,J-K_s}$ & -0.130\\
    $\rm W_{H,V-I}$ & -0.461\\
    $\rm W_{F160W,F555W-F814W}$ & -0.386\\
    \hline
    \end{tabular}
    \caption{The color term coefficients  used to calculate the Wesenheit magnitudes considered in this work: the band combinations are listed in column 1, while the $\lambda$ values, obtained by applying the Cardelli law \citep[][]{car89}, are in column 2.}
    \label{tab-wesCoeff}
\end{table}

\subsection{Monte Carlo simulations}\label{montecarlo}
In this section we present the simulations performed to test how our mathematical procedure is able to recover the correct coefficients of the fitted equation. First, we introduce the procedure adopted to simulate observational data. Then, we describe the fitting cases considered in this work.

\subsubsection{Procedure}

Following a procedure similar to that described by \citet{lay19}, we simulated 1000 samples containing the same number of stars as in the sample analyzed in this work. To simulate the observed quantities (i.e. magnitudes, ${\rm [Fe/H]}$ values and period of pulsation), we assumed that the observations reported by R21 are the characterizing  parameters (mean and standard deviation) of Gaussian distributions. Then we randomly extracted the values of magnitudes, metallicity and period from the quoted distributions. 

As for the distance, assuming a known $\rm PLZ$ or $\rm PWZ$ relations (hereafter $\rm PLZ_{true}$, $\rm PWZ_{true}$), we used the simulated data to calculate the absolute magnitude and then the 'true' parallax of all sources. 
In particular, given the pulsational period and the [Fe/H] values, the $\rm PLZ_{true}/PWZ_{true}$ relation in the generic band X allows to calculate the absolute magnitude $\rm M_X$ as:
\begin{equation}
    \rm M_X = \alpha_X + (\beta_X + \delta_X[Fe/H])\log P + \gamma_X [Fe/H]
\end{equation}
where $\rm \alpha_X$, $\rm \beta_X$, $\rm \gamma_X$ and $\rm \delta_X$, are taken from \citet[][]{rip21} for all the band combinations considered in this work.
The distance modulus is therefore given by $\rm \mu_0 = m_0 - M_X$, where $\rm m_0$ is the absorption corrected apparent magnitude, while the parallax is obtained by:
\begin{equation}
    \rm \varpi(mas) = 1000\cdot 10^{-(1+0.2\mu_0)}
\end{equation}

The observed parallax was then drawn randomly from the Gaussian distributions having as mean the 'true' parallax and as standard deviation the parallax error of the original sample. Finally, a given offset was added to the simulated parallaxes. 

To test the effect of bad data points on the fit results, we have also simulated the presence of measurement outliers. Specifically, we substituted a randomly selected fraction of the full sample absolute magnitudes, with values drawn from a uniform distribution.

Every simulated sample was then processed by the ODR routine and the obtained parameters were compared with the coefficients of the input $\rm PLZ_{true}$ or of the $\rm PWZ_{true}$ and with the chosen parallax offset. The coefficients of the $\rm PLZ_{true}$ and of the $\rm PWZ_{true}$ relations were fixed according to the recent results by R21\footnote{The choice of the coefficients used to simulate the data does not impact the results, as our procedure is aimed at verifying the ability of our algorithm to recover the coefficients themselves.}, while the parallax offset was arbitrarily chosen to be equal to 10 $\rm \mu$as, but we verified that the simulation results do not depend on the particular value of the offset.

To take into account the presence of the quoted outlier points we have performed a $\sigma$-clipping procedure by testing different values for the number of $\sigma$ ($\rm n_\sigma = 1.5, 2, 2.5, 3, 3.5, 4, 4.5, 5, 6, 7, 8, 9, 10, 15$) used to detect the bad data points.  Two outlier estimators have been considered: one based on the Median Absolute Deviation (MAD) and the other on the Double Median Absolute Deviation (DMAD), both applied to the residuals around the fit\footnote{The MAD of a distribution is defined by calculating the absolute deviations from the median and by taking their median value. It defines a symmetric width of the distribution around the median. On the other hand, the DMAD is defined by separating the absolute deviations into values below and above the median distribution and by taking the median of the two subgroups. In this way, the DMAD defines an asymmetrical width of the distribution around the median.}. 
Our first aim is to understand the effect, if any, of the chosen number of sigmas for MADs/DMADs (respectively $\rm n_{MAD}$, $\rm n_{DMAD}$) on the parameter estimation.

To quantify how much the parameters obtained from the 1000 simulations differ from the expected values, indicating with $\rm p$ the generic fitted coefficient, we calculated the quantity $\rm \Delta(\%) = 100\cdot \frac{p_{fit}-p_{true}}{p_{fit}}$. This provides an estimate of the systematic error performed by the adopted fitting routine. Together with the systematic error, we calculated also the statistical error  (hereafter $\sigma$) by considering the robust standard deviation ($\rm 1.4826\cdot MAD$) of the parameter distributions obtained by running the Monte Carlo simulations.
The obtained systematic and statistical error values are listed in Table.~\ref{tab-allCases-finalErrors}.

Hereafter we analyse the results for the following cases considered in this work: i) all the coefficients of Eq.~\ref{eq:abl-implicit} have been fitted (\abcde-case); ii) the $\delta$ coefficient has not been fitted, but its dependence is present in the adopted $\rm PLZ_{true}$ and $\rm PWZ_{true}$ relations (\abce-case); iii)  the $\delta$ coefficient has been neglected both by the fitting routine and in the adopted $\rm PLZ_{true}$ and $\rm PWZ_{true}$ relations (\abcep-case).

\subsubsection{\abcde-case}

\begin{figure}
   \centering
\includegraphics[width=8.5cm]{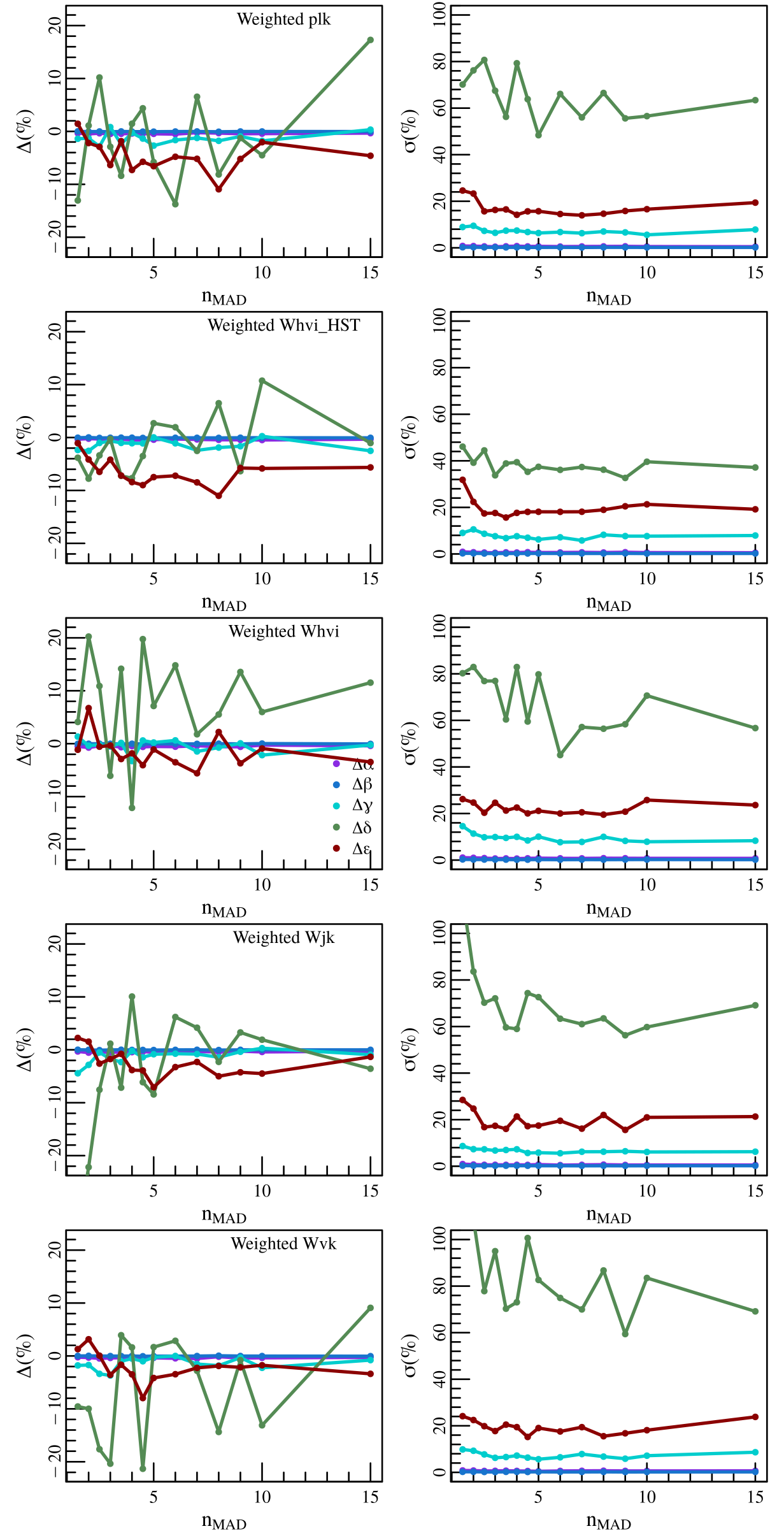}
      \caption{The results obtained by running the weighted Monte Carlo simulations for the \abcde-case. In the left panels the median systematic errors on the fitted parameters are plotted as a function of the $\rm n_\sigma$ values used for the $\sigma$-clipping procedure. The $\rm K_s$, $\rm W^{HST}_{HVI}$, $\rm W_{HVI}$, $\rm W_{JK_s}$ and the $\rm W_{VK_s}$ magnitudes are considered in the various panels from top to bottom, while the different labeled colors represent different parameters (see labels). In the right panels the corresponding relative error ($\sigma$s) on the fitted parameters are plotted against $\rm n_{MAD}$.}
      
         \label{fig-simulStatVsNsigma_mad_wt_abcde_f1o}
   \end{figure}
   
The left panels of Fig.\ref{fig-simulStatVsNsigma_mad_wt_abcde_f1o}, show the median of the $\Delta$ differences ($\Delta$s) as a function of $\rm n_{MAD}$ used for the outlier rejection. The results for the $\rm K_s$, $\rm W^{HST}_{HVI}$, $\rm W_{HVI}$, $\rm W_{JK_s}$ and $\rm W_{VK_s}$ magnitudes are plotted from top to bottom, and in each panel different colors indicate different parameters. For clarity reason, here we show only the results for the MAD case, but  for the DMAD we obtain similar conclusions.
 The right panels in Fig.~\ref{fig-simulStatVsNsigma_mad_wt_abcde_f1o} show the relative error ($\rm \sigma$s) on the fitted parameters against $\rm n_{MAD}$ for all the band combinations considered in this work. 
As a further test of the capability of the fitting routine of recovering the correct coefficients, given the difficulty to set the correct weight values in this kind of mathematical problem \citep{zwo07}, we performed a set of 1000 unweighted simulations, whose results are shown in the Fig.~\ref{fig-simulStatVsNsigma_mad_nowt_abcde_f1o} for the MAD method.

Looking at  Fig.~\ref{fig-simulStatVsNsigma_mad_wt_abcde_f1o} (\ref{fig-simulStatVsNsigma_mad_nowt_abcde_f1o}) we can assert that: i) the parameters $\alpha$ and $\beta$ are recovered with negligible both systematic (left panels) and random errors (right panels), as well as with no evident dependence on the $\rm n_\sigma$ value; ii) the $\gamma$ coefficient is not affected by significant systematic error, while the random error in many cases, amounts  to about 10-15\%; iii) the weighted (unweighted) simulations show that the $\epsilon$ coefficient is slightly underestimated (overestimated) with respect to the true value by an amount that depends on the band combination, but not larger than 10\% and with no evident dependence on the $\rm n_\sigma$; iv) the random error on $\epsilon$ always amounts to 20\% except for the cases when small $\rm n_\sigma$ values are used to reject outliers; v) the systematic error on the $\delta$ parameter has an highly variable behaviour, with large oscillations indicating both overestimation and underestimation with respect to the true value, but without a clear dependence on the $\rm n_{\sigma}$ value and/or on the photometric bands; vi) also the random error on the $\delta$ coefficient (right panel) is very large and can amount to more than 100\%.

\begin{figure}
   \centering
\includegraphics[width=8.5cm]{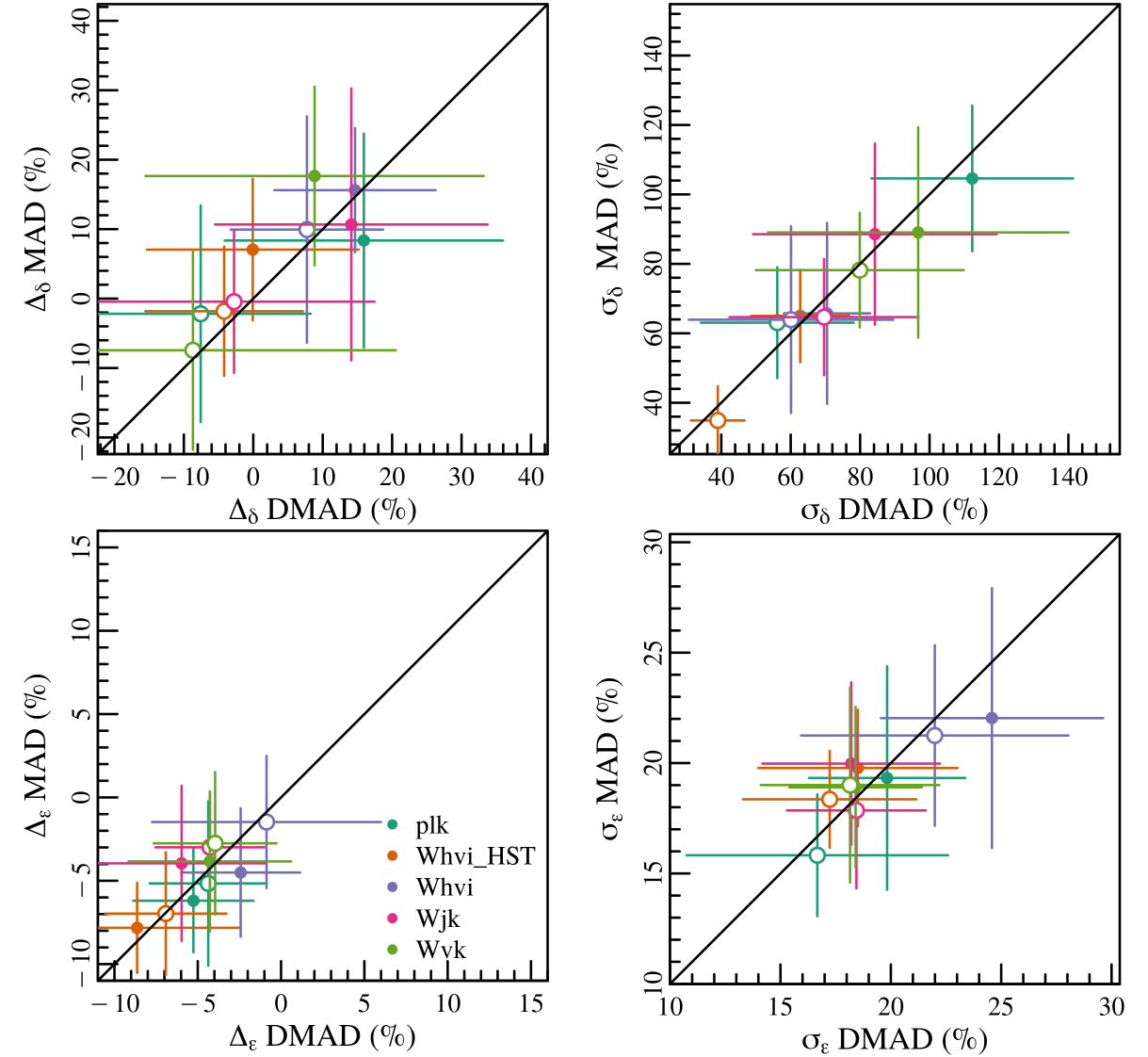}
      \caption{Summary plot of the simulation results for the $\delta$ and $\epsilon$ coefficients, including all the considered cases. The panels in the left column refer to the systematic errors, while those in the right column are for the random errors. The plotted points are the median values calculated from the simulations as described in the text, while the error bars indicate their range of variation (min-max). Different colors indicate different bands, while empty and filled symbols show the results for F and F+1O cases respectively. Moreover, the one-to-one line is also drawn to facilitate the comparison}
         \label{fig-abcde-errorsAnalysis.Wt.MadDmad}
   \end{figure}

\begin{figure}
   \centering
\includegraphics[width=8.5cm]{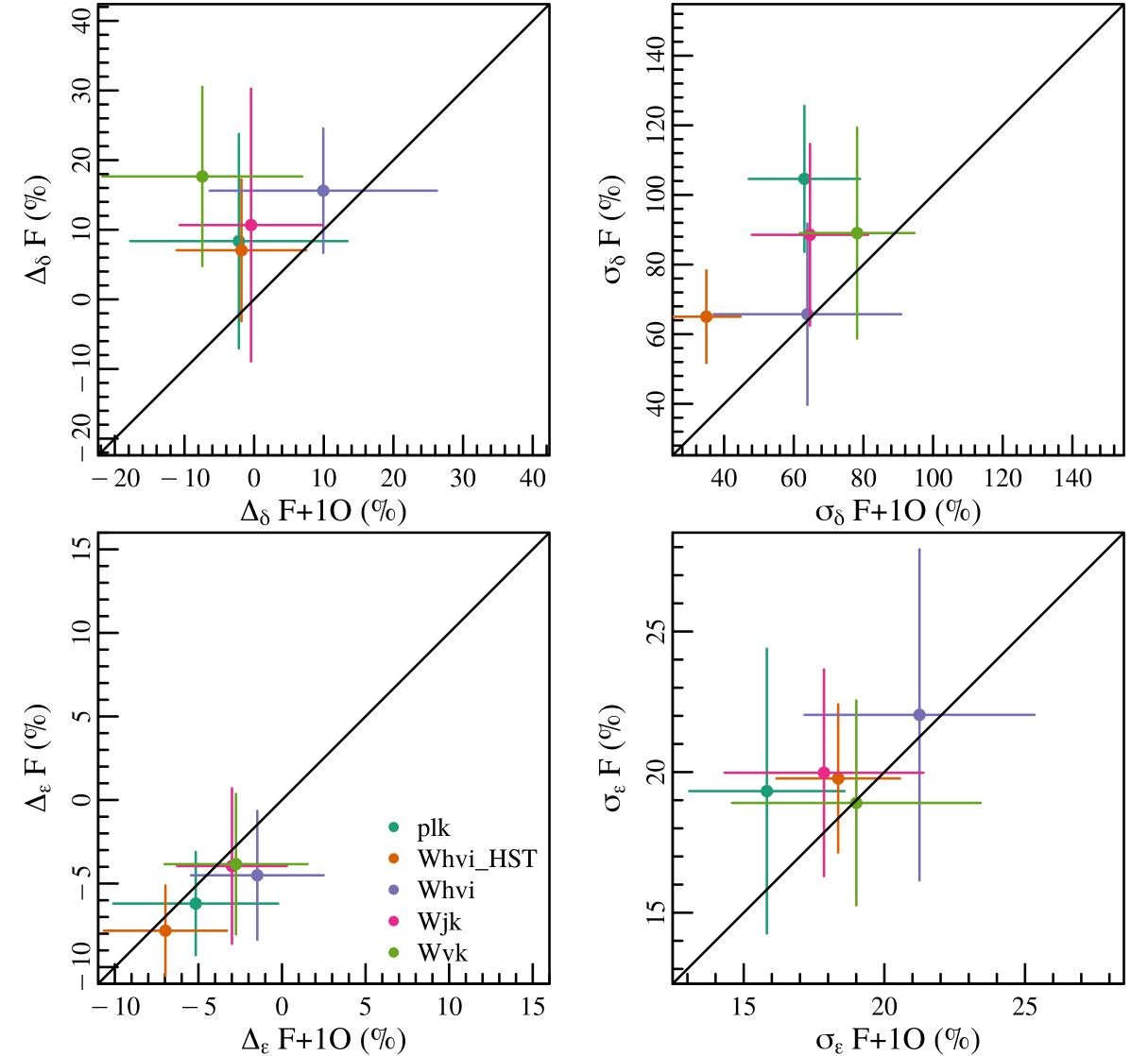}
      \caption{Summary plot of the simulation results for the $\delta$ and $\epsilon$ coefficients, including all the considered cases. The panels in the left column refer to the systematic errors, while those in the right column are for the random errors. The plotted points are the median values calculated from the simulations as described in the text, while the error bars indicate their range of variation (min-max).  Different colors indicate different bands and the one-to-one line is also drawn to facilitate the comparison. }
         \label{fig-abcde-errorsAnalysis.Wt.FOfund}
   \end{figure}

Since there is no clear dependence of the fitting results on the adopted $\rm n_\sigma$, except for very small values, we decided to define as systematic and as random errors on the fitted parameters the median of the $\Delta $s and the median of the $\sigma$s plotted in  Fig.~\ref{fig-simulStatVsNsigma_mad_wt_abcde_f1o}  (\ref{fig-simulStatVsNsigma_mad_nowt_abcde_f1o}), obtained by excluding the cases with too small $\rm n_\sigma$ values ($\rm n_\sigma$ < 2.5), subject to larger deviations. Moreover, we also calculated the min-max range covered by the quoted $\Delta$s and $\sigma$s in order to quantify their variability in the simulations. This effect is shown in Fig.~\ref{fig-abcde-errorsAnalysis.Wt.MadDmad} and Fig.~\ref{fig-abcde-errorsAnalysis.Wt.FOfund}. 
Considering that the errors on the $\alpha$, $\beta$ and $\gamma$ are negligible and the parameters are accurately recovered by the fitting routine, the reported figures focus only on the remaining  $\delta$ and $\epsilon$ coefficients.

Figure~\ref{fig-abcde-errorsAnalysis.Wt.MadDmad} shows a comparison between the outlier rejection methods. The left panels refer to the systematic errors, while the right panels to the random errors. In each panel, the results for the DMAD are plotted on the abscissa, while those for the MAD are on the ordinate. Moreover, the one-to-one line allows us to assert that both the systematic and random errors on the $\delta$ coefficient and those on the $\epsilon$ coefficient, are independent of the adopted outlier rejection method. 
Therefore, to continue our analysis we selected the results obtained from the MAD method and  compared the simulations based on the sample of F pulsators with those for the F+1O sample (see Fig.~\ref{fig-abcde-errorsAnalysis.Wt.FOfund}, in order to quantify how the inclusion of the 1O sources influences the fit results. Fig.~\ref{fig-abcde-errorsAnalysis.Wt.FOfund}  shows that the inclusion of the fundamentalized 1O stars (reported on the abscissa) into the fitted sample allows us to reduce both systematic and random errors for the $\delta$ coefficient, and mainly the random errors for what concerns $\epsilon$. On this basis, we decided to focus our analysis on the global F+1O sample.

On the basis of our Monte Carlo simulations, we decided to fit the $\delta$ parameter only for the case of the HST bands (orange point in Fig.\ref{fig-abcde-errorsAnalysis.Wt.FOfund}), characterized by a systematic error very close to zero  (top-left panel), by a contained variation ($<10\%$), and by the lowest random error, when compared with the other bands, (top-right panel) amounting to $\simeq 40\%$.

On the contrary, our routine allows a quite accurate recovery of the $\epsilon$ coefficient, with systematic errors below 10\% and random errors around 20\% for every band combinations.

The results of this analysis, in combination with those discussed in the subsequent sections,  are listed in Table~\ref{tab-allCases-finalErrors}. In particular this table contains the values of the systematic and random errors affecting all the fitted coefficients, as derived from the Monte Carlo simulations.

\subsubsection{\abce-case}

\begin{figure}
   \centering
\includegraphics[width=8.5cm]{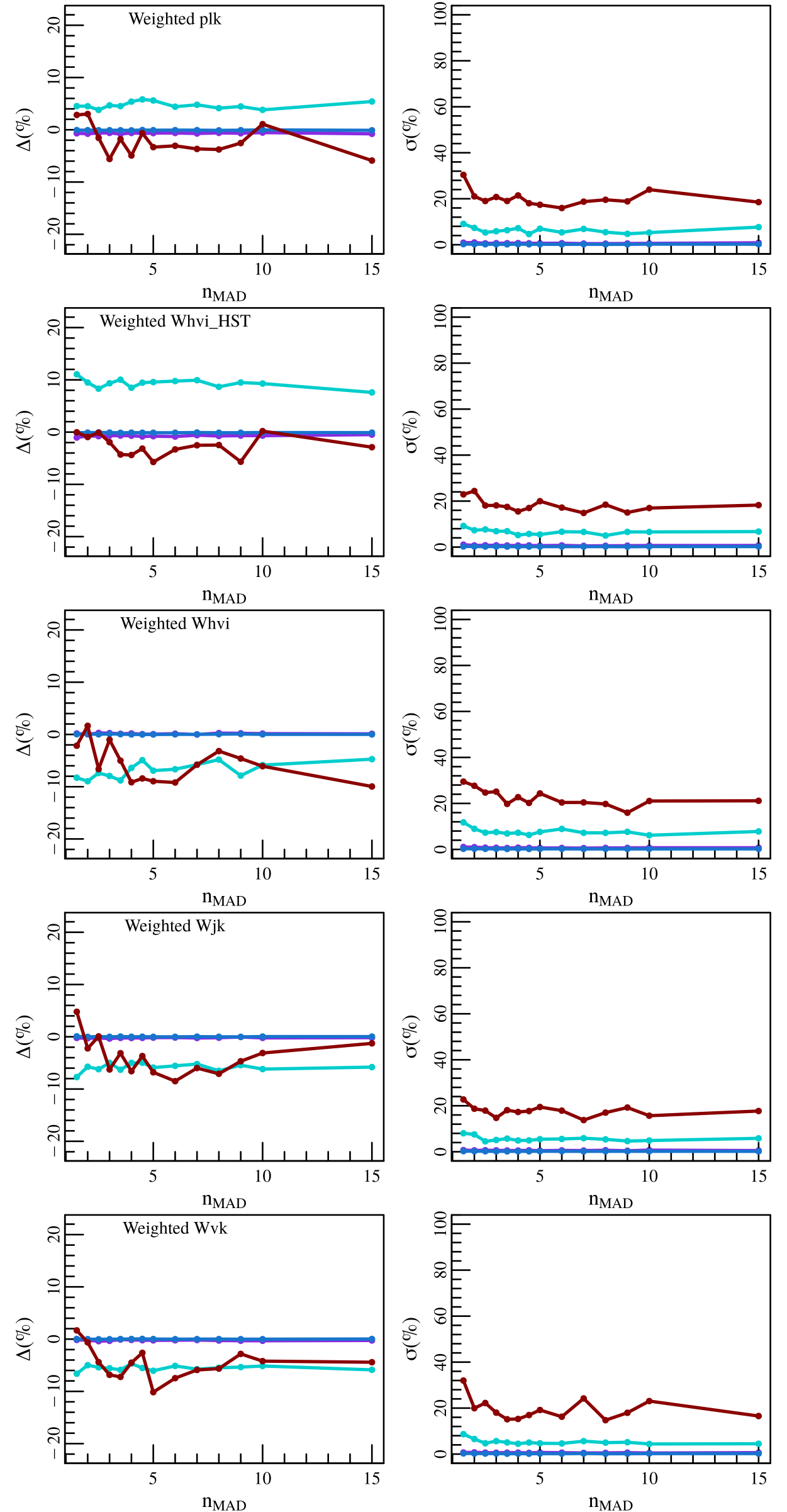}
      \caption{Results obtained by running the Monte Carlo simulations by assuming a 5-parameter $\rm PL[FeH]_{true}$ relation, but neglecting the $\delta$ coefficient in the fit. The content of the panels and the meaning of the symbols are the same as in Fig.~\ref{fig-simulStatVsNsigma_mad_wt_abcde_f1o}.}
         \label{fig-simulStatVsNsigma_mad_wt_abce_f1o}
   \end{figure}

In this case, we assume that the true $\rm PLZ$ depends on the $\delta$ parameter, but we neglect it in the fit to estimate the impact on the other coefficients of the $\rm PLZ$ relations. The results of this second Monte Carlo test are shown in  Fig.~\ref{fig-simulStatVsNsigma_mad_wt_abce_f1o} (\ref{fig-simulStatVsNsigma_mad_nowt_abce_f1o}), which is similar to Fig.~\ref{fig-simulStatVsNsigma_mad_wt_abcde_f1o} (\ref{fig-simulStatVsNsigma_mad_nowt_abcde_f1o}), but without the $\delta$ coefficient. The main difference with respect to the \abcde-case consists in the larger systematic error on the $\gamma$ parameter, which for some bands ($\rm PL_K$ and $\rm PW^{HST}_{HVI}$) is overestimated up to $\sim10\%$, and for the remaining bands is underestimated by almost the same amount. We do not show the plot for the unweighted simulations, since it implies similar results. 

The results obtained by studying this case are listed in Table~\ref{tab-allCases-finalErrors} and are labeled as \abce-wt or \abce-nowt, respectively for the weighted and the unweighted fit.

\subsubsection{\abcep-case}

The results discussed so far have been obtained by assuming the existence of a dependence on the $\delta$ coefficient in Eq.~\ref{eq:abl-implicit}. As the significance of this parameter is still uncertain (see e.g. R21), we decided to perform an additional set of simulations by assuming that this coefficient is zero in the fitted equation (hereafter \abcep case). We verified that the results for this test are similar to those shown in Fig.~\ref{fig-simulStatVsNsigma_mad_wt_abce_f1o}, with the exception of the $\gamma$ coefficient that is recovered without significant systematic errors.

As for the previously considered cases, the results for this additional set are listed in  Table~\ref{tab-allCases-finalErrors} and are labeled as \abcep$\rm _{wt}$ or \abcep$\rm _{nowt}$, respectively, for the weighted and the unweighted fit.

\begin{table*}
\centering
\begin{tabular}{lccccccccccl}
  \hline
Bands & $\Delta_{\alpha}$ & $\sigma_{\alpha}$ & $\Delta_{\beta}$ & $\sigma_{\beta}$ & $\Delta_{\gamma}$ & $\sigma_{\gamma}$ & $\Delta_{\delta}$ & $\sigma_{\delta}$ & $\Delta_{\epsilon}$ & $\sigma_{\epsilon}$ & Case \\ 
  \hline
$\rm Whvi$ & 0.47 & 0.68 & 0.07 & 0.21 & -6.1 & 7.1 & - & - & 1.9 & 23.0 & \abce$\rm _{nowt}$ \\ 
  $\rm Whvi$ & -0.17 & 0.65 & 0.02 & 0.21 & -0.27 & 7.5 & - & - & 7.4 & 24.0 & \abcep$\rm _{nowt}$ \\ 
  $\rm Whvi$ & -0.39 & 0.71 & 0.01 & 0.18 & -1.7 & 7.2 & - & - & -0.9 & 20.0 & \abcep$\rm _{wt}$ \\ 
  $\rm Whvi$ & 0.17 & 0.71 & 0.02 & 0.22 & -6.9 & 7.3 & - & - & -6.5 & 22.0 & \abce$\rm _{wt} $ \\ 
  $\rm Whvi_{HST}$ & -0.10 & 0.63 & 0.05 & 0.18 & 0.26 & 7.1 & 0.54 & 44.0 & 7.2 & 20.0 & \abcde$\rm _{nowt} $ \\ 
  $\rm Whvi_{HST}$ & -0.42 & 0.61 & -0.05 & 0.20 & -1.1 & 7.5 & -1.8 & 35.0 & -7.0 & 18.0 & \abcde$\rm _{wt} $ \\ 
  $\rm Whvi_{HST}$ & -0.49 & 0.72 & -0.01 & 0.20 & 10.0 & 6.3 & - & - & 13.0 & 21.0 & \abce$\rm _{nowt} $ \\ 
  $\rm Whvi_{HST}$ & -0.15 & 0.63 & 0.04 & 0.20 & 0.00 & 5.4 & - & - & 6.7 & 19.0 & \abcep$\rm _{nowt} $ \\ 
  $\rm Whvi_{HST}$ & -0.38 & 0.68 & -0.04 & 0.18 & -0.40 & 5.7 & - & - & -4.9 & 17.0 & \abcep$\rm _{wt} $ \\ 
  $\rm Whvi_{HST}$ & -0.69 & 0.69 & -0.10 & 0.21 & 9.3 & 6.6 & - & - & -2.6 & 18.0 & \abce$\rm _{wt} $ \\ 
  $\rm Wjk$ & 0.03 & 0.54 & 0.06 & 0.18 & -5.2 & 4.9 & - & - & 5.0 & 21.0 & \abce$\rm _{nowt} $ \\ 
  $\rm Wjk$ & -0.12 & 0.59 & 0.05 & 0.19 & -0.29 & 5.4 & - & - & 9.4 & 20.0 & \abcep$\rm _{nowt} $ \\ 
  $\rm Wjk$ & -0.39 & 0.59 & -0.03 & 0.18 & -1.0 & 5.6 & - & - & -3.2 & 19.0 & \abcep$\rm _{wt} $ \\ 
  $\rm Wjk$ & -0.22 & 0.60 & 0.01 & 0.18 & -5.7 & 5.1 & - & - & -5.3 & 17.0 & \abce$\rm _{wt} $ \\ 
  $\rm Wvk$ & 0.04 & 0.60 & 0.06 & 0.17 & -4.8 & 5.1 & - & - & 6.1 & 20.0 & \abce$\rm _{nowt} $ \\ 
  $\rm Wvk$ & -0.16 & 0.63 & 0.04 & 0.19 & -0.15 & 5.6 & - & - & 10.0 & 21.0 & \abcep$\rm _{nowt} $ \\ 
  $\rm Wvk$ & -0.36 & 0.61 & -0.03 & 0.18 & -1.3 & 5.3 & - & - & -4.1 & 19.0 & \abcep$\rm _{wt} $ \\ 
  $\rm Wvk$ & -0.27 & 0.63 & -0.01 & 0.20 & -5.5 & 4.9 & - & - & -5.1 & 18.0 & \abce$\rm _{wt} $ \\ 
  $\rm plk$ & -0.37 & 0.60 & -0.00 & 0.18 & 6.0 & 5.7 & - & - & 9.0 & 21.0 & \abce$\rm _{nowt} $ \\ 
  $\rm plk$ & -0.15 & 0.60 & 0.04 & 0.19 & -0.31 & 5.3 & - & - & 6.7 & 19.0 & \abcep$\rm _{nowt} $ \\ 
  $\rm plk$ & -0.42 & 0.60 & -0.03 & 0.18 & -1.2 & 5.1 & - & - & -3.1 & 17.0 & \abcep$\rm _{wt} $ \\ 
  $\rm plk$ & -0.62 & 0.63 & -0.08 & 0.20 & 4.6 & 5.7 & - & - & -3.2 & 19.0 & \abce$\rm _{wt} $ \\ 
  \hline
\end{tabular}
\caption{This table contains the systematic ($\Delta$s) and statistical errors ($\sigma$s), expressed in \%, used to correct the results of the ODR fit routine for all the considered cases. \abcde is the case where all 4 PLZ/PWZ coefficients are fitted and the parallax offset residual $\epsilon$ is estimated simultaneously; \abce is the case where $\delta$ is not fitted but is considered in the $\rm PLZ_{true}$ and the $\rm PWZ_{true}$ relations; finally the \abcep case is where $\delta$ is neglected both in the fitting procedure and in the $\rm PLZ_{true}$ and $\rm PWZ_{true}$ relations. The subscripts "nowt" and "wt" refer to a weighted and unweighted fitting. We note that the systematic and random errors for the $\delta$ parameter are given only for the $\rm Whvi_{HST}$ case because of too large errors for all the other band combinations, as detailed in the text.} 
\label{tab-allCases-finalErrors}
\end{table*}

\section{Results}
\label{sec-theFitResults}
In this section we discuss the coefficients derived from the PLZ/PWZ relations, together with the estimated residual parallax offset. Moreover we describe the application of our fitted relations to derive the LMC distance.

\subsection{Implicit fit of the PLZ/PWZ relations}
\label{sezione3.1}

According to the simulations described above, we carried out  the ODR procedure by considering all the cases listed in  Table~\ref{tab-allCases-finalErrors}. We focused our attention on both the weighted and the unweighted fit of the PLZ (PWZ) relations described by eq.~\ref{eq:abl-implicit}, where we neglected the $\delta$ coefficient, with the exception of the case including the HST band combination. Moreover, we included the fundamentalized 1O pulsators in the fitted sample and we did not apply any $\sigma$-clipping rejection (see Sect.~\ref{montecarlo}).
The obtained coefficients were then corrected for the systematic errors found through the simulations and listed in Table~\ref{tab-allCases-finalErrors}. In this table we give the rms of the residuals around the fitted ABL relations, even if, for comparison with other estimates, it would be easier, to have PLZ/PWZ dispersion expressed in mag. Anyway, this involves the parallax inversion to calculate the distance and consequently the observed absolute magnitude. As well-known from other studies \citep{are99, lur18, bai21}, this procedure  introduces bias in the distance estimation, that can lead to incorrect results, especially for the sources characterized by a large parallax error ($\sigma_\varpi/\varpi > 10\%$). The ABL formalism was introduced to get rid of this problem since the parallax is used linearly by definition. 
Nevertheless, we tried to give an estimate of the PLZ/PWZ scatter in mag using a non-rigorous procedure. First, for each star, we computed the absolute magnitude using the fitted PLZ/PWZ relationship, by inserting the stars’ period and [Fe/H]. Then, we calculated the observed absolute magnitudes for each star by using the apparent dereddened magnitudes (or the Wesenheit magnitudes) and the Gaia parallaxes. Finally, the scatter of the residuals between the observed and computed absolute magnitude is taken as an estimate of the dispersion of the specific PLZ/PWZ relation.  We obtained that these dispersions are similar for all the relationships, and of the order of 0.2 mag.

\begin{figure}
   \centering
   \includegraphics[width=8.5cm]{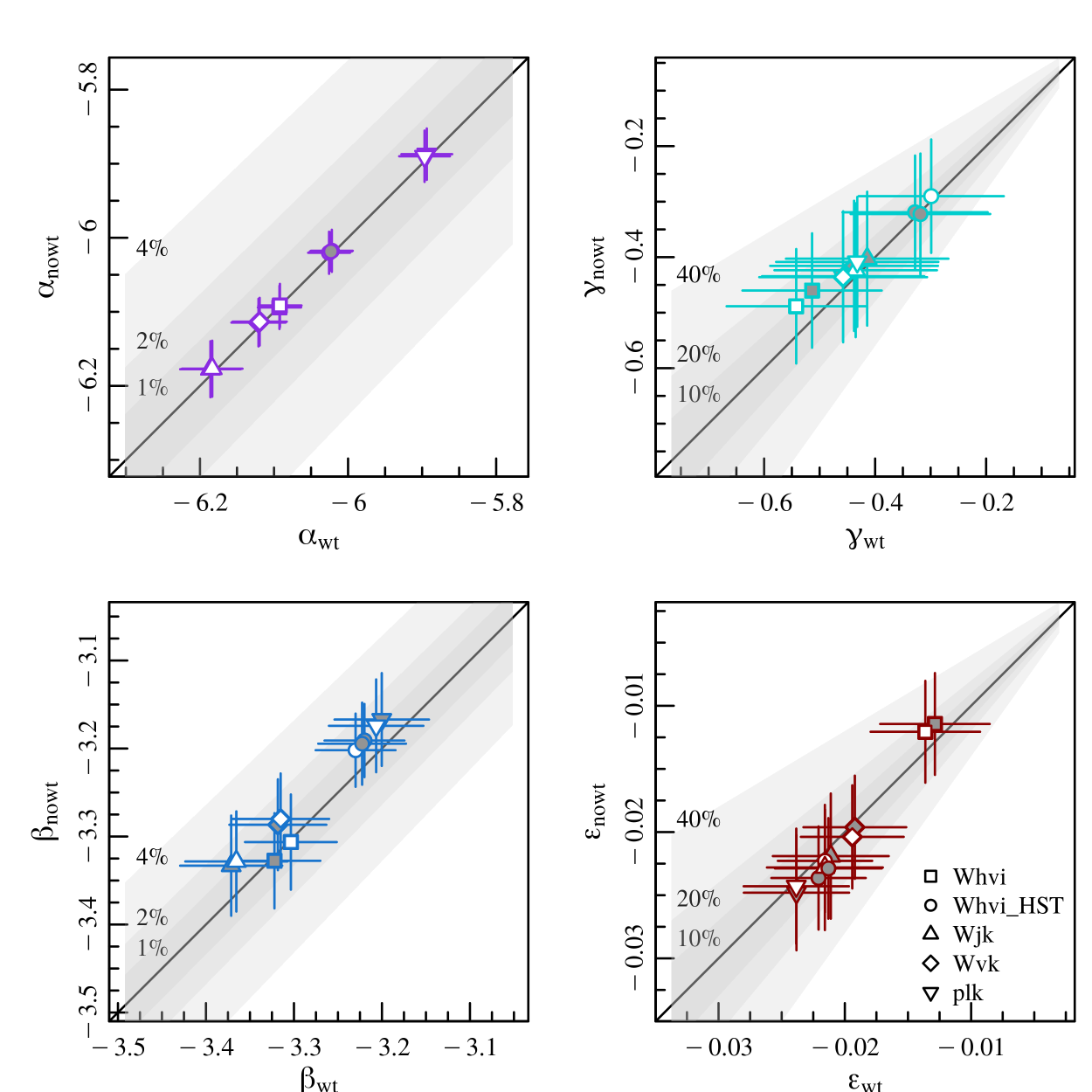}
      \caption{Comparison between the weighted and the unweighted ODR fit $\alpha$, $\beta$, $\gamma$ and $\epsilon$ coefficients listed in Table~\ref{tab-odrFit-coefficients}. The weighted values are plotted on the abscissa, while the unweighted are on the ordinates. Different symbols indicate different bands, as labeled in the bottom right panel. For a given band, gray-filled and white-filled points indicate respectively \abce and \abcep cases.
      To facilitate the comparison the one-to-one line is also shown, together with a shaded area labeled with three levels of relative difference between the plotted parameters.}
         \label{fig-odrCoeffsWtVsNowt}
   \end{figure}

The comparison of the weighted and unweighted coefficients is shown in  Fig.~\ref{fig-odrCoeffsWtVsNowt}, where the one-to-one line is drawn together with shaded areas showing three difference levels (in per cent).
The parameter best values are in good agreement, with relative differences within 2\% for $\alpha$ and $\beta$, and below 10\% for $\gamma$ and $\epsilon$. If we include the error bars, the differences between weighted and unweighted results do not rise above 4\% for $\alpha$ and $\beta$, while for the case of $\gamma$ and $\epsilon$ they can be as large as 40\%. The case of the $\delta$ coefficient is not considered in the quoted plots because it would be represented by just one point. Anyway, looking at its value in  Table~\ref{tab-odrFit-coefficients}, the weighted and the unweighted cases differ by $40\%$, even if they are not significant according to their errors. 

\begin{figure}
   \centering
   \includegraphics[width=8.5cm]{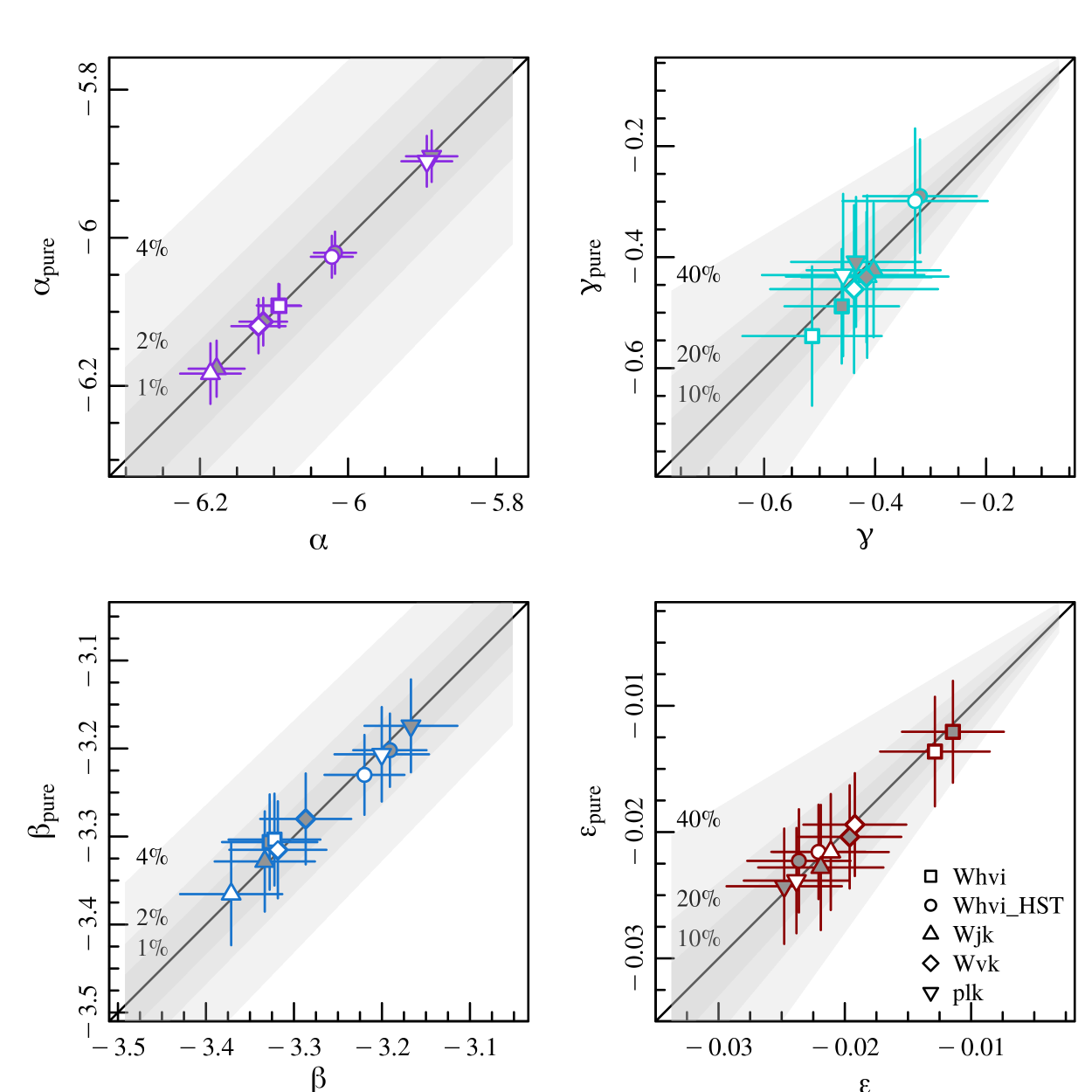}
      \caption{Same as Fig.~\ref{fig-odrCoeffsWtVsNowt} but for the comparison between the \abce and \abcep coefficients. In this case, for a given band, gray-filled and white-filled points indicate unweighted and weighted results respectively.}
         \label{fig-odrCoeffsWtVsNowt-abceVsabcePure}
   \end{figure}

The comparison between the \abce\ and \abcep cases suggests similar conclusions, with no significant difference between the obtained coefficients (see Fig.~\ref{fig-odrCoeffsWtVsNowt-abceVsabcePure}).

\begin{table*}
\centering
\begin{tabular}{lclccccccc}
  \hline
Bands & $\rm N_{dat}$ & $\rm Case$ & rms$_{\rm ABL}$ & $\alpha$ & $\beta$ & $\gamma$ & $\delta$ & $\epsilon$ & $\rm \mu_0$ \\ 
  \hline
$\rm Whvi$ & 316 & \abce$\rm _{nowt}$ & 0.017 &$-3.328 \pm 0.054$ & $-6.094 \pm 0.029$ & $-0.46 \pm 0.10$ & - & $-0.0114 \pm 0.0040$ & $18.406 \pm 0.044$ \\ 
  $\rm Whvi$ & 316 & \abcep$\rm _{nowt}$ & 0.017 &$-3.306 \pm 0.054$ & $-6.091 \pm 0.029$ & $-0.49 \pm 0.10$ & - & $-0.0121 \pm 0.0040$ & $18.401 \pm 0.054$ \\ 
  $\rm Whvi$ & 316 & \abcep$\rm _{wt}$ & 0.021 & $-3.303 \pm 0.052$ & $-6.092 \pm 0.029$ & $-0.54 \pm 0.13$ & - & $-0.0136 \pm 0.0043$ & $18.381 \pm 0.065$ \\ 
  $\rm Whvi$ & 316 & \abce$\rm _{wt}$ & 0.021 & $-3.322 \pm 0.052$ & $-6.093 \pm 0.029$ & $-0.51 \pm 0.13$ & - & $-0.0129 \pm 0.0043$ & $18.385 \pm 0.071$ \\ 
  $\rm Whvi_{HST}$ & 430 & \abcde$\rm _{nowt}$ & 0.016 & $-3.195 \pm 0.047$ & $-6.018 \pm 0.026$ & $-0.32 \pm 0.11$ & $-0.14 \pm 0.37$ & $-0.0229 \pm 0.0040$ & $18.511 \pm 0.078$ \\ 
  $\rm Whvi_{HST}$ & 430 & \abcde$\rm _{wt}$ & 0.020 & $-3.223\pm0.050$ & $-6.023\pm0.028$ & $-0.32\pm0.13$ & $-0.10\pm0.38$ & $-0.0213\pm0.0042$ & $18.499 \pm 0.078$\\ 
  $\rm Whvi_{HST}$ & 430 & \abce$\rm _{nowt}$ & 0.016 & $-3.191 \pm 0.042$ & $-6.018 \pm 0.028$ & $-0.32 \pm 0.10$ & - & $-0.0236 \pm 0.0041$ & $18.489 \pm 0.052$\\ 
  $\rm Whvi_{HST}$ & 430 & \abcep$\rm _{nowt}$ & 0.016 & $-3.202 \pm 0.042$ & $-6.020 \pm 0.028$ & $-0.29 \pm 0.10$ & - & $-0.0223 \pm 0.0041$ & $18.499 \pm0.066$\\ 
  $\rm Whvi_{HST}$ & 430 & \abcep$\rm _{wt}$ & 0.020 & $-3.230 \pm 0.045$ & $-6.026 \pm 0.028$ & $-0.30 \pm 0.13$ & - & $-0.0216 \pm 0.0037$ & $18.490 \pm 0.055$ \\ 
  $\rm Whvi_{HST}$ & 430 & \abce$\rm _{wt}$ & 0.020 & $-3.220 \pm 0.045$ & $-6.022 \pm 0.028$ & $-0.33 \pm 0.13$ & - & $-0.0221 \pm 0.0037$ & $18.478 \pm 0.059$ \\ 
  $\rm Wjk$ & 443 & \abce$\rm _{nowt}$ & 0.016 & $-3.333 \pm 0.057$ & $-6.178 \pm 0.038$ & $-0.40 \pm 0.12$ & - & $-0.0219 \pm 0.0050$ & $18.501 \pm 0.068$ \\ 
  $\rm Wjk$ & 443 & \abcep$\rm _{nowt}$ & 0.016 & $-3.328 \pm 0.057$ & $-6.177 \pm 0.038$ & $-0.42 \pm 0.12$ & - & $-0.0228 \pm 0.0050$ & $18.494 \pm 0.084$ \\ 
  $\rm Wjk$ & 443 & \abcep$\rm _{wt}$ & 0.021 & $-3.365 \pm 0.058$ & $-6.183 \pm 0.041$ & $-0.44 \pm 0.15$ & - & $-0.0216 \pm 0.0046$ & $18.477 \pm 0.064$ \\ 
  $\rm Wjk$ & 443 & \abce$\rm _{wt}$ & 0.021 & $-3.371 \pm 0.058$ & $-6.186 \pm 0.041$ & $-0.41 \pm 0.15$ & - & $-0.0211 \pm 0.0046$ & $18.485 \pm 0.091$ \\ 
  $\rm Wvk$ & 430 & \abce$\rm _{nowt}$ & 0.016 & $-3.287 \pm 0.052$ & $-6.115 \pm 0.032$ & $-0.42 \pm 0.12$ & - & $-0.0196 \pm 0.0041$ & $18.506 \pm 0.061$ \\ 
  $\rm Wvk$ & 430 & \abcep$\rm _{nowt}$ & 0.016 & $-3.280 \pm 0.052$ & $-6.113 \pm 0.032$ & $-0.44 \pm 0.12$ & - & $-0.0204 \pm 0.0041$ & $18.500 \pm 0.061$ \\ 
  $\rm Wvk$ & 430 & \abcep$\rm _{wt}$ & 0.021 & $-3.315 \pm 0.055$ & $-6.119 \pm 0.037$ & $-0.46 \pm 0.15$ & - & $-0.0194 \pm 0.0041$ & $18.479 \pm 0.072$ \\ 
  $\rm Wvk$ & 430 & \abce$\rm _{wt}$ & 0.021 & $-3.318 \pm 0.055$ & $-6.121 \pm 0.037$ & $-0.44 \pm 0.15$ & - & $-0.0192 \pm 0.0041$ & $18.487 \pm 0.070$ \\ 
  $\rm plk$ & 443 & \abce$\rm _{nowt}$ & 0.017 & $-3.167 \pm 0.053$ & $-5.887 \pm 0.035$ & $-0.43 \pm 0.12$ & - & $-0.0248 \pm 0.0046$ & $18.506 \pm 0.057$ \\ 
  $\rm plk$ & 443 & \abcep$\rm _{nowt}$ & 0.017 & $-3.174 \pm 0.053$ & $-5.890 \pm 0.035$ & $-0.41 \pm 0.12$ & - & $-0.0243 \pm 0.0046$ & $18.516 \pm 0.061$ \\ 
  $\rm plk$ & 443 & \abcep$\rm _{wt}$ & 0.022 & $-3.207 \pm 0.054$ & $-5.897 \pm 0.034$ & $-0.43 \pm 0.15$ & - & $-0.0238 \pm 0.0042$ & $18.497 \pm 0.077$ \\ 
  $\rm plk$ & 443 & \abce$\rm _{wt}$ & 0.022 & $-3.200 \pm 0.054$ & $-5.894 \pm 0.034$ & $-0.46 \pm 0.15$ & - & $-0.0238 \pm 0.0042$ & $18.486 \pm 0.061$ \\ 
   \hline
\end{tabular}
\caption{This table summarizes the results of the ODR fit: the fitted magnitude is listed in column 1; the number of fitted data points is in column 2; the flag of the fitted case defined in the text is in column 3; the rms of the residuals around the fit is contained in column 4. Note that this quantity is the rms of the ABL function and has therefore the units of the inverse of the square root of a flux \citep{are99}; columns 5-9 contain the fitted parameters; the estimated LMC distance for every considered case is contained in column 10.} 
\label{tab-odrFit-coefficients}
\end{table*}

\begin{figure}
   \centering
   \includegraphics[width=12.5cm]{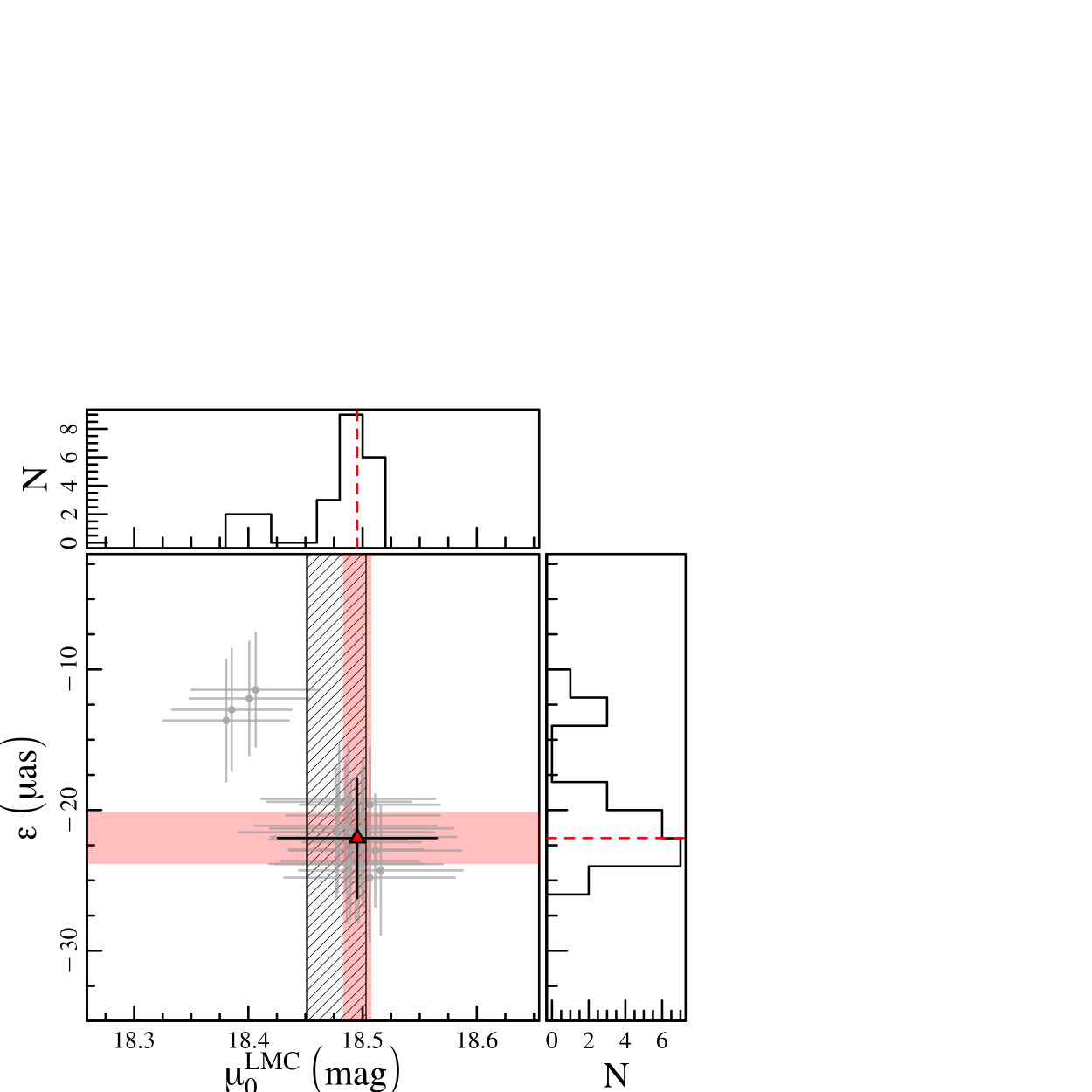}
      \caption{The central panel shows the obtained LMC distance moduli, plotted against the {\it Gaia} parallax offset values. The grey points represent the results for all the considered cases listed in column 10 of Table~\ref{tab-odrFit-coefficients}. Our best values of the two parameters are represented by the red triangle, together with the error bars defined as described in the text. The horizontal and the vertical red regions show the $1\sigma$ interval, respectively of the $\epsilon$ and $\rm \mu_0^{LMC}$ distributions plotted in the two side panels. The skewed-line region encloses the $1\sigma$ value around the currently accepted LMC distance 18.477$\pm$0.026 obtained by \citet[][]{pie19}.}
         \label{fig-parShiftVsMu0Lmc}
   \end{figure}

\begin{figure}
   \centering
   \includegraphics[width=8.5cm]{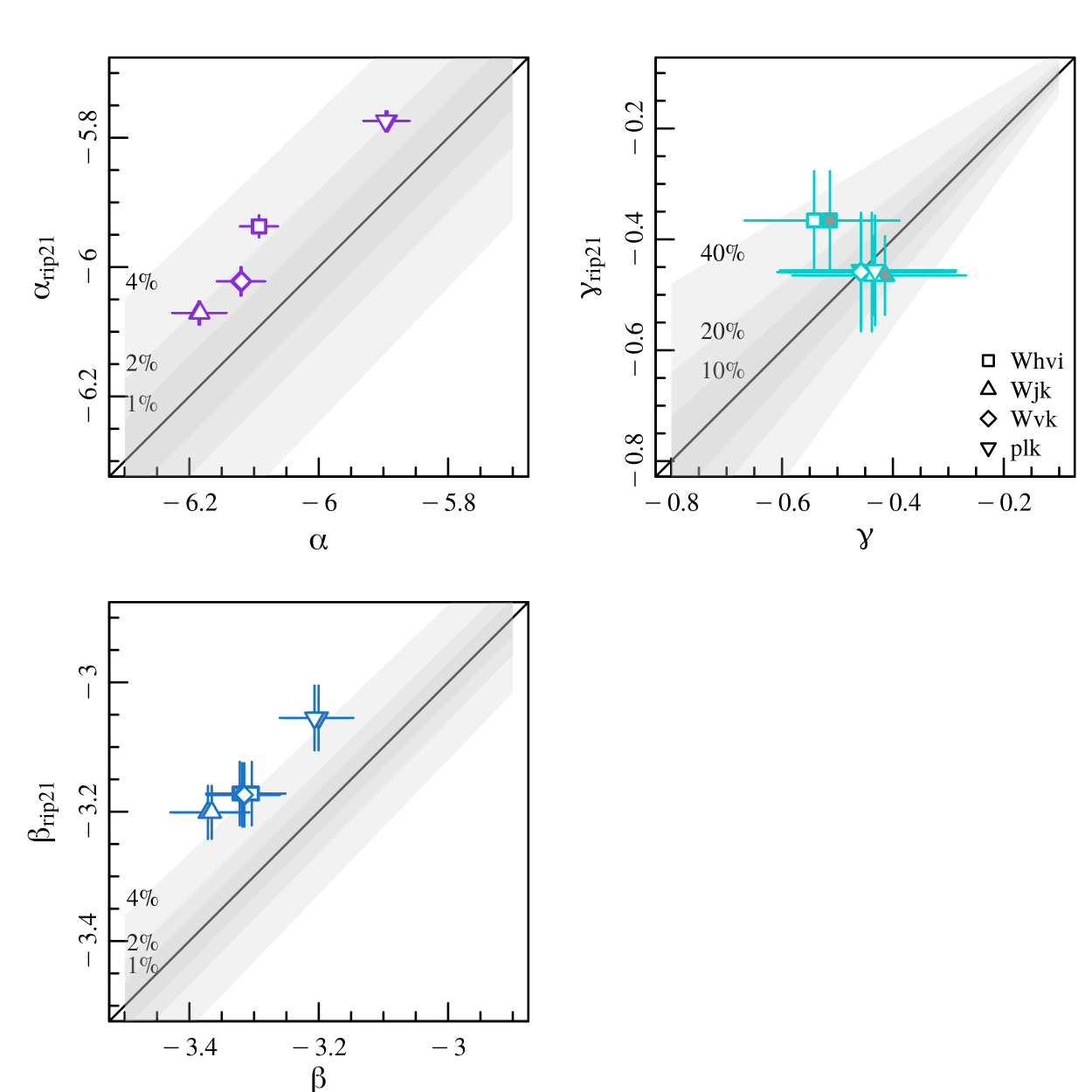}
      \caption{Same as Fig.~\ref{fig-odrCoeffsWtVsNowt} but it compares the \abce results with those by R21. Different bands are plotted with different symbols, as labeled in the top right panel. For a give band \abce and \abcep cases are plotted respectively by white-filled and gray-filled symbols.}
         \label{fig-odrCoeffsWtVsRip21}
   \end{figure}

   \begin{figure}
   \centering
   \includegraphics[width=8.5cm]{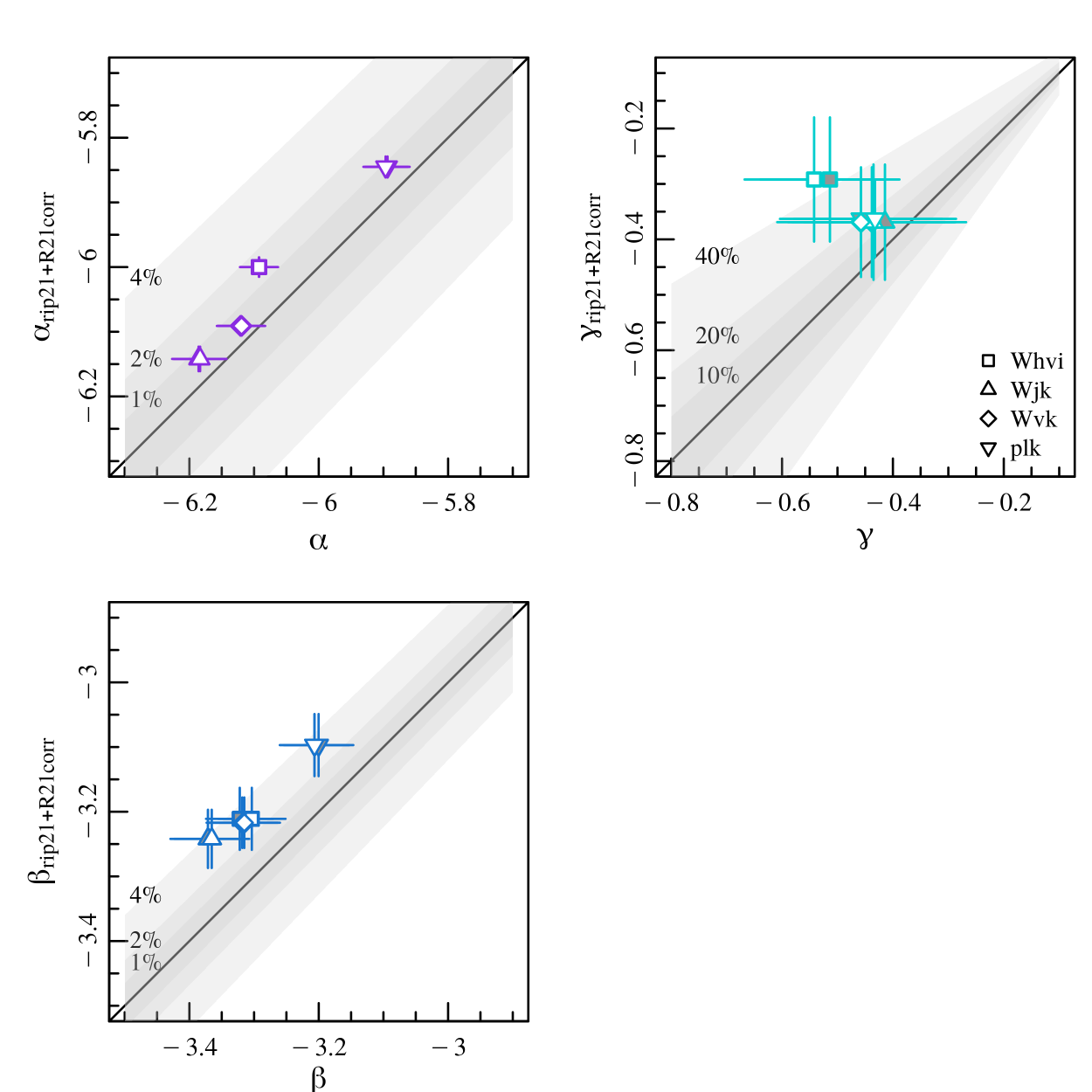}
      \caption{Same as Fig.~\ref{fig-odrCoeffsWtVsNowt} but it compares the \abce results with those by R21 including the offset by \citet[][]{rie21}. Different bands are plotted with different symbols, as labeled in the top right panel. For a give band \abce and \abcep cases are plotted respectively by white-filled and gray-filled symbols.}
         \label{fig-odrCoeffsWtVsRip21RiessOffset}
   \end{figure}
   
Looking at the values of the statistical error ($\sigma_s$) predicted by the Monte Carlo simulations and reported in Table~\ref{tab-allCases-finalErrors}, we note that they are systematically smaller than the parameter uncertainties given in Table~\ref{tab-odrFit-coefficients} which lists our final results. This is due to the fact that the latter are obtained by applying the bootstrap technique\footnote{The bootstrap simulations are obtained by re-sampling randomly the input data (allowing data point repetitions) and applying the fit to every obtained sample. The distribution of values, obtained for every fitted parameter, is then used to estimate the quoted errors.}, while the former descend from Monte Carlo simulations obtained by varying the observational properties within the data errors, as described in the previous section, and are strongly affected by possible overestimation or underestimation of the errors themselves. To be conservative, we will refer to the bootstrap results listed in Table~\ref{tab-odrFit-coefficients}.

\subsection{The distance of the LMC and the global {\it Gaia} EDR3 ZP parallax offset} 

As an application of the derived PLZ (PWZ) relations, we have calculated the LMC distance and compared it with the current accepted geometric value, $\mu_0=18.477\pm 0.026$ mag, derived from the eclipsing binaries \citep[][]{pie19}. Specifically, we applied our relations to the same LMC sample  considered by R21. This consists of about 4500 sources and the individual distances have been calculated by using all the considered PLZ relations and by assuming a mean LMC metallicity value [Fe/H]$=-0.409$ dex $(\sigma=0.076\,$ dex$)$ \citep{rom22}. The LMC distances reported in  column 10 of Table~\ref{tab-odrFit-coefficients} are obtained by taking the median of the distribution of the distances of all the selected LMC sources. To calculate the associated error, we have performed 1000 Monte Carlo simulations by varying the PLZ (PWZ) coefficients within their errors and recalculating the LMC arithmetic mean distance every time. The robust standard deviation of the obtained values represents our error.

Figure~\ref{fig-parShiftVsMu0Lmc} displays the LMC distance moduli, obtained by using all the fitted PLZ (PWZ) relations, against the parallax offset values, fitted simultaneously through the procedure described above. The main panel of this plot shows that the results from all the PLZ (PWZ) relations are tightly clustered, with the exception of those relative to the HVI band combination, arguably due to some extinction coefficient mismatch. The reported marginal histograms help to visualize the distributions of the plotted data, together with the median values (dashed red lines), adopted as our best estimates of the two parameters (excluding the discrepant HVI combination). They amount respectively to $\epsilon=-22\,\mu as$\footnote{In order to better understand the comparison with the literature, discussed in the section below, it is useful here to clarify that the negative correction value found by our fitting procedure indicates that the L21 model returns too large parallaxes.} and $\rm \mu_0^{LMC}=18.49\,mag$, and are also indicated in the main panel by the red triangle. Here the vertical and the horizontal highlighted red regions show the $\pm1\sigma$ interval of the two distributions equal to $ \pm \sim0.025$ mag and $\pm \sim3\,\mu as$ respectively for $\rm \mu_0^{LMC}$ and $\epsilon$. Anyway, in order to be conservative, the error bars associated to the point, representing our best estimates, are taken equal to the mean uncertainty of the tabulated $\rm \mu_0^{LMC}$ and $\epsilon$ values, and are equal respectively to $\pm$0.06 mag and $\pm$4$\mu as$.

The obtained LMC best distance is in excellent agreement with the accepted geometric value by \citet{pie19} and corresponds to a parallax offset value significantly different from zero, indicating that the L21 recipe applied to the EDR3 parallaxes results into over-corrected values.

\section{Discussion}
\label{sec-discussion}
In this section we compare the coefficients of the obtained PLZ/PWZ relations with those obtained from a variety of literature sources.

\subsection{Comparison with other PLZ/PWZ coefficients}
We have compared the ODR fit results with those obtained recently by R21 by using the same sample of this work but calculated by means of a standard non-linear fitting approach. R21 carried out the PLZ/PWZ relations fit adopting two choices for the global corrections to the individual ZP parallax offsets by L21, namely no global correction at all and a correction of $-14\pm6~ \mu$as \citep[][]{rie21}.  

The comparison for the first case is shown in Fig.~\ref{fig-odrCoeffsWtVsRip21} for the parameters $\alpha$, $\beta$ and $\gamma$. %where the ODR values are plotted on the abscissa, while the coefficients by \citet{rip21} are on the ordinate. 
The differences between our results and R21 for the $\alpha$ and $\beta$ coefficients are within 5\%, even if, given the small uncertainties, not consistent with each other at more than 1$\sigma$ level. Conversely, the $\gamma$ parameters are consistent with those by R21.
The same comparison for the case of global ZP parallax offset as in \citet[][]{rie21} is shown in Fig.~\ref{fig-odrCoeffsWtVsRip21RiessOffset}. The $\alpha$ and $\beta$ coefficients are now consistent within $\sim~1~sigma$, while the $\gamma$ coefficients found here are larger (in absolute value), even if consistent within 1$\sigma$, compared with R21.   

We can use the distance of the LMC to further compare present results with those by R21. They found that the adoption of the global parallax ZP offset has two effects: i) it lowers the absolute value of $\gamma$ and ii) it provides a distance for the LMC larger than the reference geometric measurement. In this context, the adoption of the global parallax ZP offset found here, namely $-22~\mu$as, would imply an even larger LMC distance if we had to recalculate the PLZ/PWZ relations by using the standard non-implicit fitting method as in R21.      
A possible interpretation of this occurrence is the existence of a degeneration (or correlation) between the different coefficients of the fit and the global ZP offset. In the case of the standard non-implicit fitting method, it seems that the largest correlation is with the $\gamma$ coefficient, while in the case of the implicit method presented here, the $\alpha$ and $\beta$ values are more affected than $\gamma$. 
This somewhat unexpected behaviour can be tentatively explained taking into account the very different fitting techniques and weighting schemes adopted here and in R21. We expect that the adoption of a larger DCEP sample including e.g. a wider range in metallicity, which should be available in the near future, will allow us to better understand how to break the correlation/degeneracy between the different coefficients of the PLZ/PWZ relations.   

We note that, in the C-MetaLL project, we take advantage of high-resolution spectroscopy and multi-band photometry coupled with the {\it Gaia} parallaxes for a wide sample of Galactic DCEPs to fit simultaneously all the coefficients of the PLZ/PWZ relations, thus including the metallicity dependence. On the contrary, in the recent literature the metallicity dependence is tackled by adopting a double-step method: i) The PL/PW relations (no $\gamma$ in the Eq.~\ref{eq:abl}) are fitted to separate DCEP samples characterized by different mean Z values (e.g. DCEPs belonging to different galaxies); ii) a linear fit is performed to study the dependence of the obtained PL/PW intercepts on the Z mean value of each sample. 
Albeit this work is mainly aimed at demonstrating the capability of the implicit fit method, in the following we briefly discuss some of the DCEP PLZ/PWZ relations recently presented in the literature.

\citet{Gieren2018} adopted the Baade-Wesselink technique to derive multi-band PL/PW relations  ($\rm Mag = \alpha + \beta(\log P - log P_0)$) for three samples consisting of 32 Small Magellanic Cloud (SMC), 22 LMC and 14 MW (Milky Way) DCEPs, respectively. According to their results the $\beta$ coefficient does not change significantly for the three analyzed galaxies, supporting the hypothesis of a metal independent  PLZ/PWZ slope. To study the metallicity dependence of the PL/PW zero point, they adopted a mean metallicity value for each considered galaxy and found that the $\gamma$ coefficient ranges between -0.221$\pm$0.053 mag/dex and -0.335$\pm$0.059 mag/dex, depending on the photometric bands.

In their recent comprehensive work, \citet{Riess2022} have estimated the metallicity dependence of the PW relation in the HST bands using DCEPs in each geometric anchor galaxy (MW, LMC, SMC, NGC\,4258), thus using a method similar to \citet{Gieren2018}.  In the end they obtained  $\gamma=-0.217\pm0.046$ mag/dex.

The PLZ/PWZ relations  in 15 filters from mid-IR to optical wavelengths were studied by \citet{bre22}. They obtained these relations  for three samples of DCEPs belonging to MW, LMC and SMC. According to their results there is no strong evidence to reject the hypothesis of a metal independent slope $\beta$. Fixing the slope to the LMC value, they studied the intercept dependence on the metallicity. If we exclude their smallest and largest values, obtained in the {\it Gaia} bands, their $\gamma$ coefficient range between -0.20 and -0.33 mag/dex.

The PLZ/PWZ relations were studied also from the theoretical point of view. \citet{des22} constructed a large set of DCEP pulsational models, characterized by different metallicity values from sub-solar to super-solar, and fitted the PLZ/PWZ relations in the {\it Gaia}, HST-WFC and Johnson-Cousin photometric systems. They obtained negative $\gamma$ coefficient, as in the observational studies, albeit with a slightly smaller values.

\subsection{Comparison with other ZP parallax offsets}
Several authors studied possible residual parallax offset after the application of the correction  provided by L21. Different distance tracers, including e.g. quasars, eclipsing binaries, variable stars, open clusters and red clump stars, were used to compare the EDR3 parallaxes to independent distance estimations. 
According to L21, the correction zero-point for the {\it Gaia} EDR3 parallaxes is known to be dependent on the ecliptic latitude, the color and the G-band magnitude of the sources. Therefore a direct comparison of our best estimate with those from other authors is not always meaningful. 

\citet{fab21} compared the EDR3 parallaxes with the distances obtained from other external catalogues (see their Tab.1). According to their results, the L21 correction significantly improves the parallax comparison with the exception of the sources belonging the LMC, SMC and two dwarf spheroidal galaxies (Fornax and Sculptor).

\citet{hua21} studied the residual offset after L21 correction by using a sample of 65000 red clump stars. According to their results the L21 correction allows to reduce the parallax bias from about -26$\rm \mu as$, before correction, to  -4$\rm \mu as$.

A slight residual offset is also found by \citet{vas21}, who compared the distances of 170 MW globular clusters, collected from the literature, with those released within EDR3. According to their results, the L21 recipe leaves a residual offset equal to -10$\pm$3 $\rm \mu as$ in the sense that the corrected parallaxes are larger than the reference distances.

\citet[][]{wan22} used a sample of 0.3 million giant stars with a distance known better than $\sim 10 \%$ to find that the official L21 model largely reduces the parallax bias, with a residual offset amounting to $\rm +2.6\,\mu as\,(+2.9\,\mu as)$ for the five-parameter (six-parameter) solutions, in the sense that the corrected {\it Gaia} parallaxes are slightly bigger than the reference giant star values.

Similar conclusions were derived by \citet{ren21}, who used a sample of 0.11 million W-Ursae Major eclipsing binaries (EB) to find a residual parallax offset equal to $\rm +4.2 \pm 0.5\, \mu as\, (+4.6\pm 3.7\, \mu as) $, for the 5-parameters (6-parameters) solutions after the application of the L21 correction.

Using a sample of 158 eclipsing binaries with known bolometric luminosity,
 \citet[][]{sta21} found a difference between the L21 corrected {\it Gaia} parallaxes and those of the reference sample ($\rm EDR3^{corr} - EB$) equal to $\rm -15\pm18\,\mu as$, even if they consider this difference not statistically significant according to the reported uncertainty. 
\citet[][]{bha21} used a new calibrated PL relation in the Near Infrared bands to obtain the distances for a sample of Galactic RR Lyrae stars and compared them with those from EDR3 release. They found that the parallaxes obtained by applying the L21 model are over-corrected by $\rm +22\pm2 \mu as$ in the sense that the L21 parallaxes are too large, in excellent agreement with our findings.

A sample of RR Lyrae variables together with the same implicit fitting procedure of this work, were considered by \citet[][]{gil21}. They calibrated the PLZ relations in the Wide-field Infrared Survey Explorer (WISE) W1 and W2 bands and simultaneously fitted the {\it Gaia} EDR3 parallax offset. Their results indicate that the offset value amounts to $\rm +10\pm7\,\mu as$ or $\rm -20\pm6\,\mu as$, respectively by including or excluding the L21 correction.

Using a sample of 74 Galactic DCEPs with Hubble Space Telescope photometry  \citet{rie21} fitted simultaneously the PWZ and the residual parallax offset, finding that the L21 model over-corrects the raw EDR3 parallaxes by $\rm +14\pm6\,\mu as$ (value already used in this work), in the sense that the corrected values are too big, in agreement with our conclusions and those from the works cited above. 

A similar result was reported also by \citet{zin21}, who analyzed the asteroseismic data of a sample of red giants observed by Kepler, finding that the corrected EDR3 parallaxes are larger than the reference ones by $\rm +15\pm5\,\mu as$.

In a very recent result by \citet[][]{cru22}, the authors estimated the {\it Gaia} parallax offset by considering both the non-variable open cluster members and the open cluster Cepheids. Their main results are the following: i) the use of non variable cluster stars returns an offset equal to $\rm -4\pm 6\, \mu as$, hinting that the L21 model adequately corrects the EDR3 parallaxes in the magnitude range 12.5 < G < 17 mag; ii) the comparison of the LMC distance, estimated by their calibrated Leawitt laws (obtained in different optical and near-infrared band combinations), with the currently accepted geometric value by \citet[][]{pie19}, indicates a residual parallax offset value ranging between $\rm -17\pm 5\,\mu as$ and $\rm -22\pm 3\,\mu as$, in perfect agreement with the results reported in this work.

\begin{figure}
   \centering
   \includegraphics[width=8.5cm]{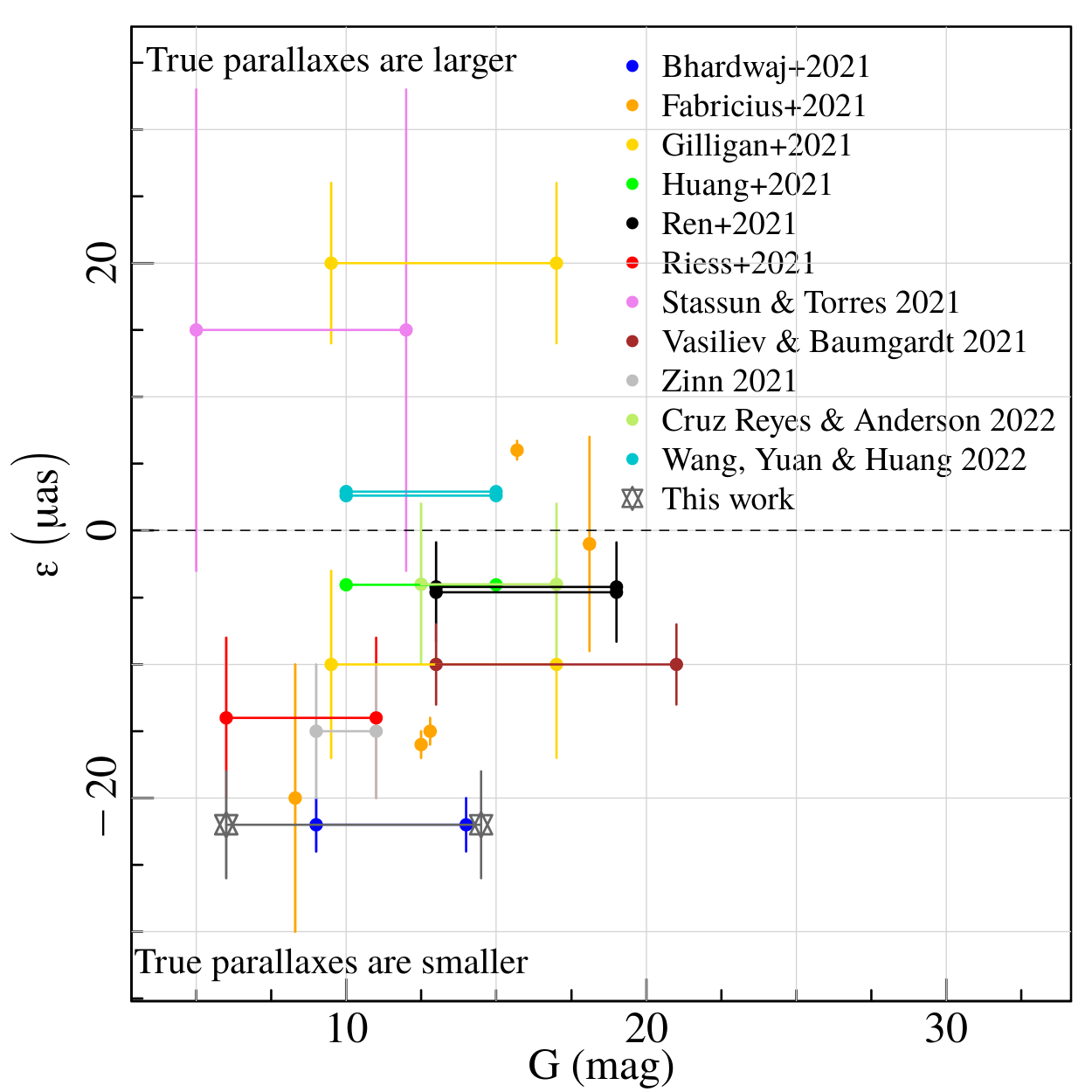}
      \caption{The offset values from the labeled references are plotted as function of the G magnitude range considered in each work. Our estimate is plotted with an empty star.  The sign of the $\epsilon$ values are such that negative (positive) offset indicate that the true reference parallaxes are smaller (larger) than the EDR3 ones after the L21 correction.}
         \label{fig-offsetComparison}
   \end{figure}

To facilitate the comparison of the reported offset values we show Fig.~\ref{fig-offsetComparison} similar to those from \citet{Riess2022} and \citet{li22}. The difference between the reference parallaxes and the EDR3 values is on the ordinate, while the G band magnitude for each considered reference sample is on the abscissa. The offset values discussed above are plotted with different colors, while our result is represented by the empty gray star. In the quoted figure, negative (positive) offset values indicate that the true reference parallaxes are smaller (larger) than those obtained after the application of the L21 correction.

\section{Conclusions}\label{sec-conclusions}
In this work we derived simultaneously the coefficients of the DCEPs PLZ (PWZ) relations and the expected offset for the {\it Gaia} parallaxes. To this aim an implicit form of the fitted relations was adopted, including both the coefficients $\alpha$, $\beta$, $\gamma$ and $\delta$, characterizing the PLZ (PWZ) relation, together with the parallax offset $\epsilon$. The solution of the implicit fit was achieved by using the ODRPACK95 FORTRAN package, which performs both unweighted fit and weighted fit, taking into account the presence of errors on all the considered variables.

A set of 1000 Monte Carlo simulations, constructed by assuming a fiducial PLZ (PWZ) relation and considering a systematic shift for the {\it Gaia} parallaxes,  was performed in order to test the capability of this tool to recover both reliable coefficient values and the given offset. The presence of possible outlier measurements was also simulated in order to study how the fitting results were affected by these bad points.

As a result of our Monte Carlo experiment, we found that: i) the presence of bad measures does not affect significantly the correct recovery of the PLZ (PWZ) relations; ii) the best results are obtained by enlarging the fitted sample through the inclusion of the fundamentalized 1O pulsators; iii) an important finding of our simulations is that the \abce case yields larger systematics for the $\gamma$ parameter than the \abcde case, while in the \abcep case the dependence on the metallicity is recovered without significant systematics; iv) the determination of metallicity coefficient $\delta$ was affected by large errors, both systematic and statistic and we decided not to derive it. This occurrence confirms that the metallicity dependence on the slope is difficult to be constrained, in agreement witth our previous results in R21.

We applied the implicit fit technique to a sample of DCEPs already used by R21, obtaining simultaneously the coefficients of a variety of PLZ (PWZ) relations as well as the relative parallax offset to be applied to the L21 corrections.  
According to our findings, our best estimate of the parallax offset ($\rm \epsilon = -22\pm 4\,\mu as$) almost cancels the L21 correction, thus indicating that the quoted model over-correct the EDR3 {\it Gaia} parallaxes, in agreement with the recent findings by other authors.
As a further test, these results, including the parallax offset, were used to calculate the distance to the LMC. The obtained value, $\rm \mu_0=18.49\pm \, 0.06\, mag$, is in perfect agreement with the current accepted geometric distance.

This work, albeit not definitive, demonstrated the validity of the implicit fit technique to estimate simultaneously both the PLZ (PWZ) relations and the parallax offset and will be applied to forthcoming works of our group,  adopting an improved DCEPs sample possessing precise photometry and accurate chemical abundances.

\section*{Acknowledgements}
The authors thank the Referee for the useful comments that made our work more readable and robust.

This work has made use of data from the European Space Agency (ESA) mission
{\it Gaia} (\url{https://www.cosmos.esa.int/gaia}), processed by the {\it Gaia}
Data Processing and Analysis Consortium (DPAC,
\url{https://www.cosmos.esa.int/web/gaia/dpac/consortium}).
Funding for the DPAC has been provided by national institutions, in particular the institutions participating in the {\it Gaia} Multilateral Agreement.
In particular, the Italian participation
in DPAC has been supported by Istituto Nazionale di Astrofisica
(INAF) and the Agenzia Spaziale Italiana (ASI) through grants I/037/08/0,
I/058/10/0, 2014-025-R.0, and 2014-025-R.1.2015 to INAF (PI M.G. Lattanzi).

\section*{Data Availability}
All the data used in this work have already been published and can be found in \citet[][and references therein]{rip21}.

%%%%%%%%%%%%%%%%%%%% REFERENCES %%%%%%%%%%%%%%%%%%

% The best way to enter references is to use BibTeX:

%\bibliographystyle{mnras}
%\bibliography{example} % if your bibtex file is called example.bib

% Alternatively you could enter them by hand, like this:
% This method is tedious and prone to error if you have lots of references

%%%%%%%%%%%%%%%%%%%%%%%%%%%%%%%%%%%%%%%%%%%%%%%%%%

%%%%%%%%%%%%%%%%% APPENDICES %%%%%%%%%%%%%%%%%%%%%

\appendix

\section{Monte Carlo simulations}
\subsection{The 5-parameter model: unweighted fit of $\alpha$, $\beta$, $\gamma$, $\delta$ and $\epsilon$}\label{app:monteCarlo-5Params}
   \begin{figure}
   \centering
\includegraphics[width=8.5cm]{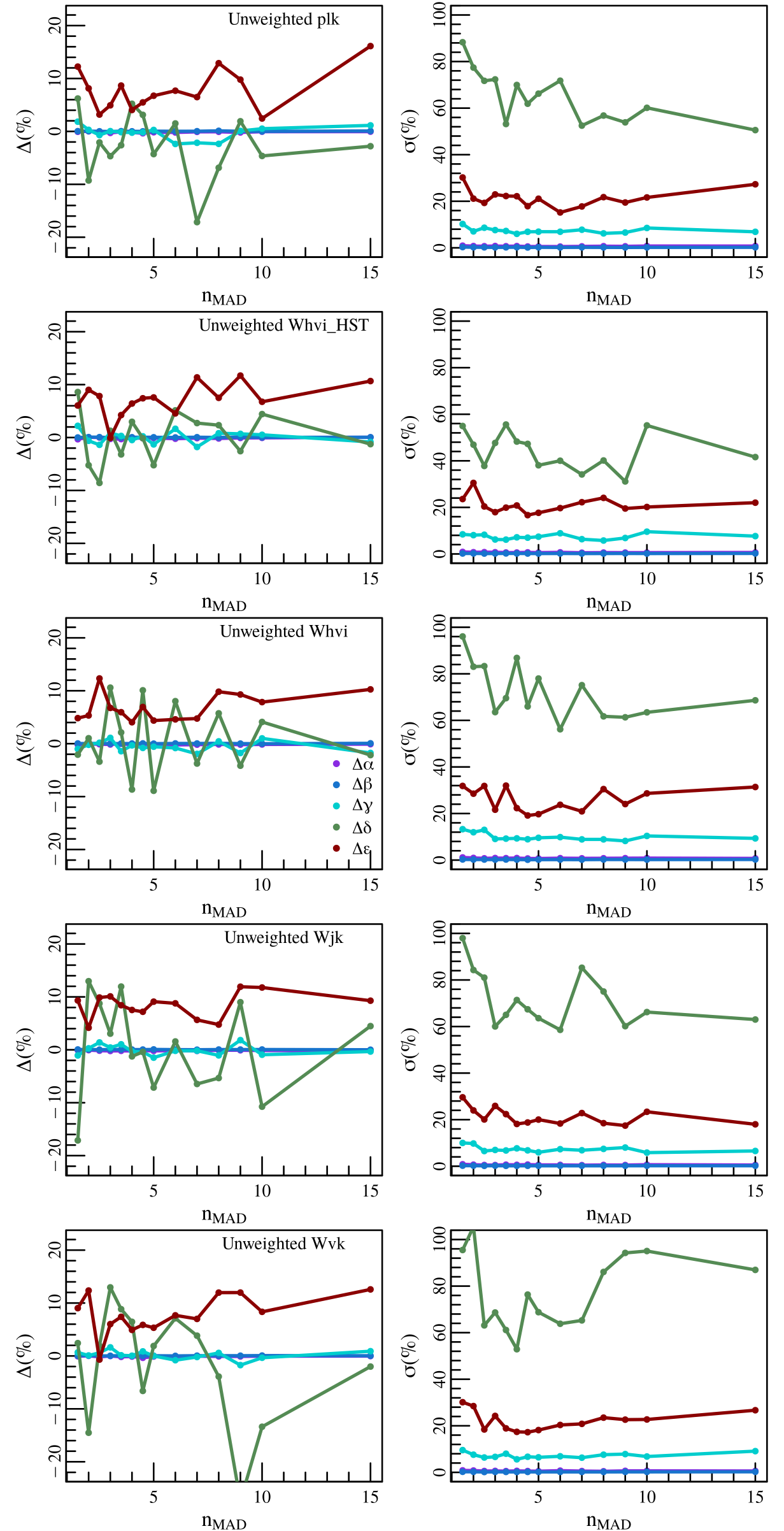}
      \caption{This figure shows the results obtained by running the Monte Carlo simulations by assuming a 5-parameters PL[FeH] relation and performing an unweighted fit. The meaning of the panels and of the symbols are the same as in the Fig.~\ref{fig-simulStatVsNsigma_mad_wt_abcde_f1o}.}
         \label{fig-simulStatVsNsigma_mad_nowt_abcde_f1o}
   \end{figure}

\subsection{The 4-parameter model: neglecting $\delta$}\label{app:monteCarlo-4Params}
\begin{figure}
   \centering
\includegraphics[width=8.5cm]{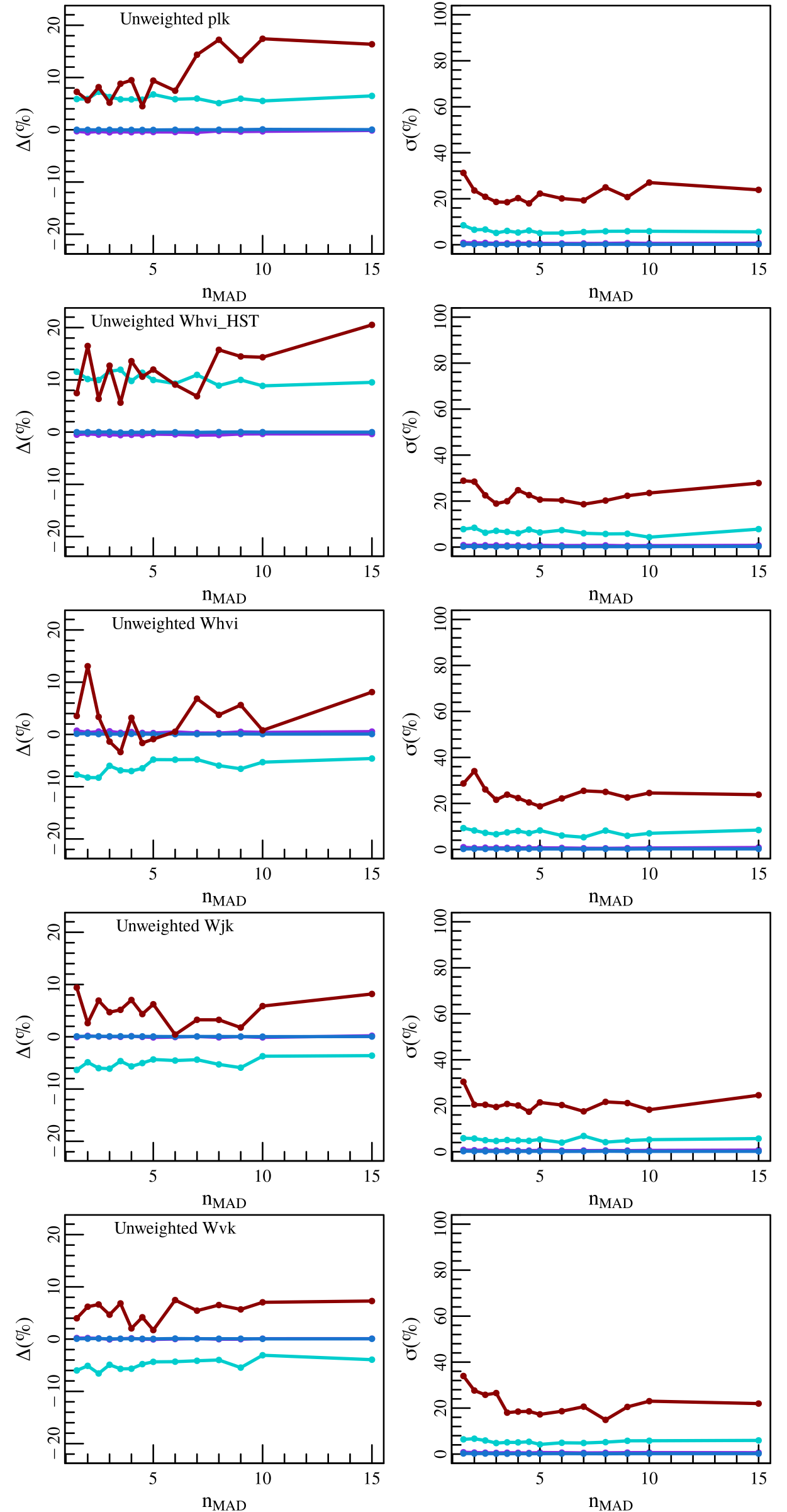}
      \caption{This figure is the same as ~\ref{fig-simulStatVsNsigma_mad_nowt_abcde_f1o} but obtained by neglecting the dependence on $\delta$ in both the true and the fitted PLZ (PWZ) relations.}
         \label{fig-simulStatVsNsigma_mad_nowt_abce_f1o}
   \end{figure}

% Don't change these lines
\bsp	% typesetting comment
\label{lastpage}
\end{document}